\documentclass[final,3p,times]{elsarticle}

\biboptions{sort&compress}

\usepackage{color}
\usepackage{amssymb}
\usepackage{amsmath}
\usepackage{mathptmx}
\usepackage{bm}
\usepackage{mathrsfs}
\usepackage{footnote}

\journal{Nuclear Physics A}

\begin{document}

\begin{frontmatter}



\title{Phi meson spectral moments and \\
QCD condensates in nuclear matter}


\author[label1]{Philipp Gubler}
\ead{gubler@ectstar.eu}

\author[label2]{Wolfram Weise}
\ead{weise@tum.de}

\address[label1]{ECT*, Villa Tambosi, 38123 Villazzano (Trento), Italy}
\address[label2]{Physik-Department, Technische Universit\"at M\"unchen, 85747 Garching, Germany}

\begin{abstract}
A detailed analysis of the lowest two moments of the $\phi$ meson spectral function in vacuum and nuclear matter is 
performed. The consistency is examined between 
the constraints derived from finite energy QCD sum rules and the spectra computed within an improved vector dominance model, 
incorporating the coupling of kaonic degrees of freedom with the bare $\phi$ meson. 
In the vacuum, recent accurate measurements of the $e^+ e^- \to K^+ K^-$ cross section allow us to determine 
the spectral function with high precision. In nuclear matter, the modification of the 
spectral function can be described by the interactions of the kaons from $\phi \rightarrow K\overline{K}$ with the surrounding nuclear medium. This leads primarily 
to a strong broadening and an asymmetric deformation of the $\phi$ meson peak structure. 
We confirm that, both in vacuum and nuclear matter, the zeroth and first moments of the corresponding spectral functions satisfy the 
requirements of the finite energy sum rules to a remarkable degree of accuracy. Limits on the strangeness sigma term of the nucleon are examined in this context.
Applying our results to the second moment of the spectrum, we furthermore discuss 
constraints on four-quark condensates and the validity of the commonly used ground state saturation approximation. 
\end{abstract}





\end{frontmatter}


\section{\label{Intro} Introduction}
Exploring the properties of the $\phi$ meson at finite baryon density is an interesting and important topic, both from a theoretical and 
experimental point of view \cite{Hayano,Leupold}. Theoretically, works based on the QCD sum rule approach \cite{Hatsuda,Klingl,Koike,Zschocke} have established a linear relation between the in-medium mass shift of the $\phi$ meson and the (scalar) strangeness content of the nucleon: $\langle N| \overline{s} s |N \rangle$ (for recent work along this line, see \cite{Gubler}). 
These studies have, however, not been able to realistically consider the width of the $\phi$. Calculations based on 
$SU(3)$ meson-baryon effective field theory using kaonic degrees of freedom predicted an increase of the $\phi$ meson width at normal nuclear matter density by about an order of 
magnitude compared to the vacuum width \cite{Klingl3,Cabrera}. 
From an experimental point of view, the study of a possible modification of the $\phi$ meson in nuclear matter has also  attracted 
considerable interest, as a number of dedicated experiments intending to measure its in-medium width and its possible  mass shift have 
either been carried out in recent years \cite{Muto,Sakuma,Polyanskiy} or will be performed in the near future at the J-PARC facility \cite{Kawama,Aoki}. 
For the interpretation of the various experimental findings, a thorough theoretical understanding of the $\phi$ meson spectral function at finite density is mandatory. 

This paper is a more detailed and extended follow-up of our earlier work \cite{GublerWeise}. We focus on the lowest three moments 
of the $\phi$ meson spectral function in vacuum and nuclear matter. The operator product expansion and the finite energy sum 
rules (FESR) relate these moments directly to the QCD condensates of lowest mass dimension and their respective in-medium modifications. 
Specifically, the $n$-th spectral moment is related to condensates with mass dimension $2(n+1)$. The advantage of first considering only the 
lowest two moments $(n = 0,1)$ lies in the fact that only condensates of up to mass dimension four are needed as an input. These are 
the well-known gluon and chiral condensates. Their behavior in nuclear matter is also relatively well understood, 
at least to leading (linear) order in the baryon density. Further subleading twist-2 corrections that appear only at finite density can be related 
to moments of quark and gluon distributions in the nucleon. 
For the second $(n = 2)$ moment of the $\phi$ meson spectral function, much less constrained structures show up in the 
operator product expansion (OPE). Among those the four-quark condensates are most prominent. Order-of-magnitude estimates for these four-quark condensates can be 
made using a factorization assumption, the quantitative reliability of which is however doubtful and will be examined in the present work. 

The strategy pursued in the present paper is briefly summarized as follows. First we construct the spectral function of the $\phi$ meson channel in the vacuum using an 
improved vector dominance model that takes into account the coupling of the $\phi$ meson to kaon pairs, introducing 
energy dependent self-energy terms in the $\phi$ propagator. This allows for an accurate 
description of the $e^+ e^- \to K^+ K^-$ cross section in the region of the $\phi$ meson peak, 
measured by the BABAR Collaboration \cite{Lees}. 
By computing the zeroth and first moment of this spectral function, its consistency with the FESR can be checked. 
This FESR is derived in QCD from an OPE of the vector current involving strange quarks. 
%
Next, this analysis is extended to nuclear matter where both the spectral function and QCD condensates are modified 
independently. To describe the in-medium $\phi$ meson spectrum, we take into account the self-energy corrections 
due to interactions of the decay kaons with nucleons in the surrounding medium. At leading order in the baryon density these corrections 
are described by the free forward $KN$ and $\overline{K}N$ scattering amplitudes for which we employ results derived from 
chiral $SU(3)$ effective field theory. For the $\overline{K}N$ channel this includes $\overline{K}N \leftrightarrow \pi \Sigma$ 
coupled channels and the non-perturbative generation of the $\Lambda(1405)$ resonance \cite{Ikeda,Ikeda2}. 
On the other hand, the density dependence of the condensates that appear on the OPE side of the FESR 
are, at leading order in $\rho$, derived model-independently using the Feynman-Hellman theorem. 
With these ingredients, 
the consistency of the lowest two moments of our in-medium spectral functions with the FESR is again tested. 
For the first moment of the spectral function in nuclear matter, its relationship to the strangeness sigma term of the nucleon is of special interest in 
view of recent lattice QCD results.

In order to estimate systematic uncertainties of our approach we examine the possible dependence of our findings on the high-energy 
properties of the spectral function. Here an issue is the modeling of the onset of the perturbative QCD continuum. In the standard FESR this onset is 
parametrized using a simple step-function. We generalize this crude description using a schematic ramp-function with a variable slope parameter 
and study its influence on the results both in vacuum and nuclear matter. 
Furthermore, we study the second moments of our vacuum and in-medium spectral functions and discus constraints on the four-quark condensates appearing in the OPE for the strange vector channel. 

This paper is organized as follows. Section \ref{Vacuum} presents our moment analysis of the vacuum $\phi$ meson spectral function 
including a discussion of the improved vector dominance model, the fit to the experimental $e^+ e^- \to K^+ K^-$ data and the matching with FESR. 
Section \ref{Nuclmatter} deals with the generalization of this approach to nuclear matter, first discussing the computation of the 
in-medium spectrum based on the $KN$ and $\overline{K}N$ scattering amplitudes and then repeating the spectral moment analysis of the previous section. Section \ref{Sigma} presents an 
updated study of the relationship between the first moment of the in-medium $\phi$ spectrum and the strangeness sigma term of the nucleon.
In Section \ref{Fourquark} the four-quark condensate values extracted from the second moment of the spectral function are presented. Finally, a summary and conclusions follow in Section \ref{Conclusion}. 

\section{\label{Vacuum} Spectral moment analysis in vacuum}
\subsection{\label{Phen} The vacuum spectral function}
The starting point is the correlator of the strange quark component of the electromagnetic current, $j_{\mu}(x) = -\frac{1}{3} \overline{s}(x) \gamma_{\mu} s(x)$, 
which couples to the physical $\phi$ meson state: 
\begin{equation}
\Pi_{\mu\nu}(q) = i\displaystyle \int d^{4}x \,\,e^{iqx} \langle \mathrm{T} [j_{\mu}(x) j_{\nu}(0)] \rangle_{\rho}. 
\label{eq:veccorr1}
\end{equation}
Here $\langle \, \rangle_{\rho}$ stands for the expectation value with respect to 
the ground state of nuclear matter at temperature $T=0$ and baryon density $\rho$. The vacuum case is realized in the limit $\rho = 0$. The polarization tensor
$\Pi_{\mu\nu}(q)$ can generally be decomposed into longitudinal and transverse components \cite{Klingl}. 
For a $\phi$ meson at rest with respect to the nuclear medium, these two components coincide and it therefore 
suffices to study the (dimensionless) contracted correlator,  
\begin{equation}
\Pi(q^2) = \frac{1}{3q^2} \Pi^{\mu}_{\mu}(q). 
\label{eq:veccorr2}
\end{equation}
Using an improved vector dominance model \cite{Klingl2}, $\mathrm{Im}\,\Pi(q^2)$ can be 
written as: 
\begin{equation}
\mathrm{Im}\Pi(q^2) = \frac{\mathrm{Im}\,\Pi_{\phi}(q^2)}{q^2 g_{\phi}^2} 
\Bigg| \frac{(1-a_{\phi})q^2 - \mathring{m}_{\phi}^2}{q^2 - \mathring{m}_{\phi}^2 - \Pi_{\phi}(q^2)} \Bigg|^2~,
\label{eq:veccorr3}
\end{equation}
where $\mathring{m}_{\phi}$ denotes a ``bare" $\phi$ meson mass. 
The coupling of the $\phi$ to $K\overline{K}$ loops and their propagation determines the self-energy $\Pi_{\phi}(q^2)$ (of dimension $mass^2$) 
in vacuum and in the nuclear medium. 
The bare mass $\mathring{m}_{\phi}$ and the coupling strength $g_{\phi}$ are fitted to experimental data in vacuum. This 
coupling strength is expected to be of the order of the 
$SU(3)$ value of $g_{\phi} \simeq -3g/\sqrt{2}$, with $g=6.5$. 
The constant $a_{\phi}$ represents the ratio between the $\phi K \overline{K}$ and $\phi \gamma$ couplings 
and should be close to unity \cite{Klingl,Klingl2}. Here we assume $a_{\phi}=1$, 
which corresponds to the limit of exact vector meson dominance where all the photon-hadron interaction is carried by vector mesons. 
This leads to a good fit to recent experimental $e^+ e^- \to K^+ K^-$ data as will be demonstrated below. 
Contributions from both charged and neutral kaon loops are included in the $\phi$ self-energy:\footnote{The subleading term describing the decay of the $\phi$ meson into three pions is ignored here.} 
\begin{equation}
\Pi_{\phi}(q^2) = \Pi_{\phi \to K^+ K^-}(q^2) + \Pi_{\phi \to K^0_L K^0_S}(q^2). 
\label{eq:self.energy}
\end{equation}
For the $K^+ K^-$ term we have \cite{Klingl2,Klingl3}: 
\begin{equation}
\Pi_{\phi \to K^+ K^-}(q^2) = \frac{-ig^2}{6} \int \frac{d^4 p}{(2 \pi)^4} 
\Big[\frac{(2p-q)^2}{(p^2 - m_{K^{\pm}}^2 + i \epsilon) ((p-q)^2 - m_{K^{\pm}}^2 + i \epsilon)} - \frac{8}{p^2 - m_{K^{\pm}}^2 + i \epsilon} \Big]. 
\end{equation}
The first term under the integral describes a genuine $K^+ K^-$ loop while the second term stands for a tadpole contribution. Evaluating first the imaginary part of this integral and then 
computing the real part using a once-subtracted dispersion relation, one obtains 
\begin{align}
\mathrm{Im}\,\Pi_{\phi \to K^+ K^-}(q^2) =&\, -\frac{g^2}{96 \pi} q^2 \Big(1 - \frac{4m_{K^{\pm}}^2}{q^2} \Big)^{3/2} \Theta(q^2 - 4m_{K^{\pm}}^2), \\
\mathrm{Re}\,\Pi_{\phi \to K^+ K^-}(q^2) =&\; b_0\, q^2 - \frac{g^2}{48 \pi^2} \Big[q^2 \mathcal{G}(q^2, m_{K^{\pm}}^2) -  4m_{K^{\pm}}^2\Big]~. 
\label{eq:RePi}
\end{align}
The function $\mathcal{G}(q^2, m^2)$ is defined as 
\begin{align}
\mathcal{G}(q^2, m^2) = 
\begin{cases}
\Big(\frac{4m^2}{q^2} -1  \Big)^{3/2} \arcsin \Big(\frac{\sqrt{q^2}}{2m}\Big)
\qquad & (0 < q^2 < 4 m^2), 
\\
-\frac{1}{2} \Big(1 - \frac{4m_{K^{\pm}}^2}{q^2} \Big)^{3/2} \ln \Big( \frac{1 + \sqrt{1-4m^2/q^2}}{1 - \sqrt{1-4m^2/q^2}} \Big)
\qquad & (4m^2<q^2,\, q^2<0).
\end{cases}
\end{align}
Following \cite{Klingl2}, the subtraction constant appearing in Eq.\,(\ref{eq:RePi}) is fixed as $b_0 = 0.11$. The corresponding 
$\overline{K}^0 K^0$ contribution in Eq.\,(\ref{eq:self.energy}) can be obtained simply replacing $m_{K^{\pm}}$ by $m_{K^0}$ in 
the above equations. 

The remaining parameters to be determined in Eq.\,(\ref{eq:veccorr3}) are $\mathring{m}_{\phi}$ and $g_{\phi}$. 
Their values are fixed by fitting Eq.\,(\ref{eq:veccorr3}) 
to the recent precise measurement of the $e^+ e^- \to K^+ K^-$ cross section
provided by the BABAR Collaboration \cite{Lees}. 
Since only the charged kaons are detected in this reaction, the corresponding $\phi \rightarrow K^+K^-$ term of  
$\mathrm{Im}\,\Pi_{\phi}(q^2)$ in the numerator of Eq.\,(\ref{eq:veccorr3}) has to be retained while 
intermediate charge exchange processes, $K^+K^-\leftrightarrow K^0\overline{K}^0$, are included in the
resummation of the $K\overline{K}$ loops. 
To describe the data at energy regions in the continuum above the $\phi$ meson peak for which the 
simple model of Eq.\,(\ref{eq:veccorr3}) is not sufficient, we add a  
second order polynomial in $c(q^2) = \sqrt{q^2/q^2_{\mathrm{th}}-1}$, 
for $\sqrt{q^2} > \sqrt{q^2_{\mathrm{th}}}=1040 ~\mathrm{MeV}$: 
\begin{equation}
\mathrm{Im}\Pi^{\mathrm{cont.}}(q^2) = A\,c(q^2) + B\,c^2(q^2),  
\label{eq:continuum}
\end{equation}
with coefficients $A$ and $B$ fitted to the data. 
We will leave this 
form of the $K^+ K^-$ continuum unchanged for both vacuum and nuclear matter cases, 
as there is presently no reliable information on the behavior of the continuum in nuclear matter. 

\begin{figure}
\begin{center}
\includegraphics[width=10cm]{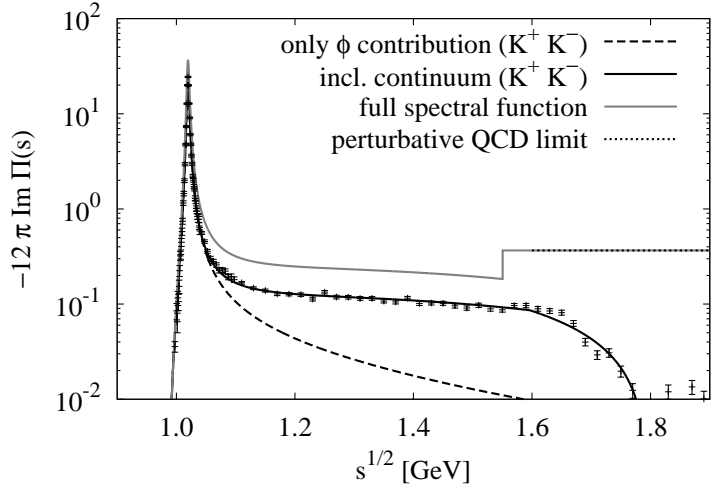}
\end{center}
\vspace{-0.6cm}
\caption{The fitted spectral function $-12 \pi \mathrm{Im} \Pi(s = q^2)$ in vacuum, compared to the experimental data 
for $\sigma(e^+e^- \to K^+ K^-)/\sigma(e^+e^- \to \mu^+ \mu^-)$, adapted from \cite{Lees}. 
The dashed [solid] curve shows the result when only Eq.(\ref{eq:veccorr3}) [both Eqs.\,(\ref{eq:veccorr3}) and (\ref{eq:continuum})] are used for the fit. 
The dotted horizontal line stands for the perturbative QCD limit while the gray line represents the full spectral function of Eq.\,(\ref{eq:ansatz}), 
including $K^0\overline{K}^0$ and other channels.}
\label{fig:vac.fit}
\end{figure}

Performing the fit one finds $g_{\phi}=0.74 \times (-3g/\sqrt{2}) \simeq -10.2$, 
$\mathring{m}_{\phi} = 797\,\, \mathrm{MeV}$, $A=-5.94 \times 10^{-3}$ and $B=3.61 \times 10^{-3}$. 
The resulting curve is shown in Fig.\,\ref{fig:vac.fit} together with the experimental $e^+ e^- \to K^+ K^-$ data. 
As demonstrated in this figure, the parametrizations (\ref{eq:veccorr3},\ref{eq:continuum}) 
give an accurate description of the data up to about $\sqrt{s} \simeq 1.6~\mathrm{GeV}$, above  
which the experimental points for $e^+ e^-\rightarrow K^+ K^-$ are seen to drop rapidly. 
This drop is parametrized by a simple linear curve fitted to the data points in this region. 

On first sight the behavior of the spectral function just described would appear to be in contradiction 
with Eq.(\ref{eq:ansatz}) of the next section which approaches the perturbative QCD limit at high energy. 
However, the $K^+K^-$ data alone do not give the complete picture of the spectral function probed by the FESR, which 
includes all possible channels that can couple to the strange vector current. 
These are, in particular, channels with two kaons and one or more pions, which 
start to give a sizable contribution to the spectrum at energies 
around and above $1.6$ GeV, as will be discussed in more detail in Section\,\ref{RampSection}. 
The behavior of the  full spectral function including all relevant channels is shown 
schematically by the gray line in Fig.\,\ref{fig:vac.fit}. 

\subsection{\label{Sumrulesvac} Finite-energy sum rules}
The correlator of Eqs.\,(\ref{eq:veccorr1}) and (\ref{eq:veccorr2}) in 
the deep-Euclidean region ($Q^2 = -q^2 \to \infty$)
can be expanded with the help of the operator product expansion. In the vacuum one has: 
\begin{equation}
9\,\Pi(q^2=-Q^2) =  -c_0 \log \Big(\frac{Q^2}{\mu^2}\Big) + \frac{c_2}{Q^2} + \frac{c_4}{Q^4} + \frac{c_6}{Q^6} +  \dots.   
\label{eq:ope1}
\end{equation}
The coefficients $c_i$ are\footnote{The $\lambda_a$ in $c_6$ denote Gell-Mann SU(3) color matrices.}
\begin{align} 
c_0 =& \frac{1}{4 \pi^2}\Big(1 + \frac{\alpha_s}{\pi} \Big), \hspace{1cm}
c_2 = -\frac{3 m_s^2}{2 \pi^2}, \label{eq:operesult1} \\
c_4 =& \frac{1}{12} \Big \langle \frac{\alpha_s}{\pi} G^2 \Big \rangle + 2m_s \langle \overline{s} s \rangle, \label{eq:c4} \\
c_6 =& -2 \pi \alpha_s \Big[ \langle (\overline{s}\,\gamma_{\mu} \gamma_5 \,\lambda^a\,s)^2 \rangle  + \frac{2}{9} 
         \langle (\overline{s}\,\gamma_{\mu} \,\lambda^a \,s) \sum_{q=u,d,s} (\overline{q}\,\gamma_{\mu} \,\lambda^a \,q)  \rangle \Big] 
         + \frac{m^2_s}{3}\Big[{1\over 3} \Big\langle \frac{\alpha_s}{\pi} G^2 \Big \rangle -8 m_s \langle \overline{s} s \rangle\Big]~.   
\label{eq:operesult3}
\end{align}
Further terms of higher order in $\alpha_s$ and $m_s$ have also been computed \cite{Gubler,Generalis,Chetyrkin,Loladze,Surguladze}. We retain here 
only the most important contributions, sufficient for the purposes of the present work. 

Applying the Borel transform to the once subtracted dispersion relation 
\begin{equation}
\Pi(q^2) = \Pi(0) + \frac{q^2}{\pi}  \displaystyle \int_0^{\infty} ds \frac{\mathrm{Im}\Pi(s)}{s(s - q^2 -i\epsilon)},  
\label{eq:displ}
\end{equation}
one can derive the sum rule: 
\begin{equation}
\frac{1}{M^2} \int_0^{\infty} ds \, R(s) \, e^{-s/M^2} = c_0 + \frac{c_2}{M^2} + \frac{c_4}{M^4} + \frac{c_6}{2M^6} +  \dots
\label{eq:sumrule}
\end{equation}
with the spectral function 
\begin{equation}
R(s) = -\frac{9}{\pi} \mathrm{Im}\, \Pi(s).
\end{equation}
For sufficiently large $s$ this spectral function approaches the perturbative QCD limit, $c_0$. We therefore introduce the following simple ansatz: 
\begin{equation}
R(s) = R_{\phi}(s)\, \Theta(s_0 - s) + R_{\mathrm{c}}(s) \,\Theta(s - s_0), 
\label{eq:ansatz}
\end{equation}
with $R_{\mathrm{c}}(s) = c_0$.  Here $s_0$ stands for a scale that delineates the low-energy and high-energy parts of the spectrum. 
After substituting into Eq.\,(\ref{eq:sumrule}) and expanding the left-hand side in inverse powers of $M^2$, 
the finite-energy sum rules are derived as 
\begin{align}
\int_0^{s_0} ds \, R_{\phi}(s) &= c_0 \,s_0 + c_2, \label{eq:sm1} \\
\int_0^{s_0} ds \, s \, R_{\phi}(s) &= \frac{c_0}{2} s_0^2 - c_4, \label{eq:sm2} \\
\int_0^{s_0} ds \, s^2 R_{\phi}(s) &= \frac{c_0}{3} s_0^3 + c_6. \label{eq:sm3}
\end{align}
In this work we focus primarily on the first two moments (\ref{eq:sm1},\ref{eq:sm2}) for which the Wilson 
coefficients and corresponding operator expectation values are known with relatively good accuracy. 

\subsection{\label{RampSection} Improved modeling of the continuum: a schematic ramp function}
\begin{figure}
\begin{center}
\hspace*{-0.4cm}
\includegraphics[width=12cm]{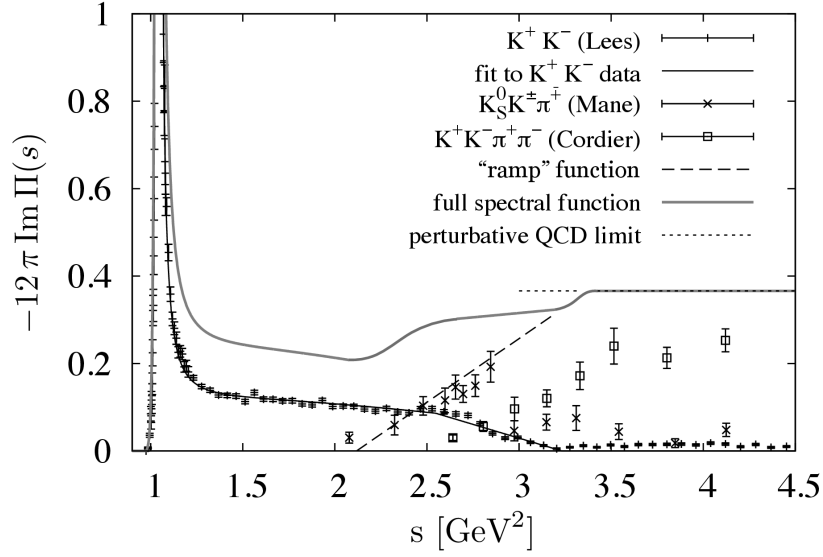}
\end{center}
\vspace{-0.8cm}
\caption{The spectral function $-12 \pi \mathrm{Im} \Pi(s)$ in vacuum, with $K^+ K^-$ data in the 
low energy region \cite{Lees} and additional data for several $2K + n\pi$ channels above $s \simeq 1.9\,\mathrm{GeV}^2$ also shown \cite{Cordier,Mane}. 
A fit to the $K^+ K^-$ data is presented together with a (linear) ramp function (dashed) and the full interpolating spectral function that reaches asymptotically the 
perturbative QCD limit (dotted).}
\label{fig:vac.fit.2}
\end{figure}

One could raise the question whether the step function adapted in Eq.(\ref{eq:ansatz}) to describe the onset of 
the QCD continuum is sufficiently accurate for analyzing the sum rules and whether a more elaborate parametrization would 
lead to qualitatively different results. To examine this possibility we introduce the following 
extended ansatz using a ramp function $W(s)$ \cite{Kwon}: 
\begin{equation}
R(s) = R_{\phi}(s) \Theta(s_2 - s) + R_{\mathrm{c}}(s) W(s,s_1,s_2), 
\label{eq:ansatz.ramp}
\end{equation}
with $W(s,s_1,s_2)$ defined as 
\begin{equation}
W(s,s_1,s_2) = 
\begin{cases}
0  & (s<s_1)
\\
\frac{s - s_1}{s_2 - s_1} & (s_1 \leq s \leq s_2)
\\
1 & (s>s_2)
\end{cases}. 
\label{eq:ansatz.ramp2}
\end{equation} 
This form is motivated by experimental data in the intermediate region above 
the $\phi$ meson resonance where states with one or two pions in addition to $K\overline{K}$ start to play a significant role. 
A selection of such data (the $e^+ e^- \to K_S^0 K^{\pm} \pi^{\mp}$ and $K^+ K^- \pi^+ \pi^-$ channels) are shown in 
Fig. \ref{fig:vac.fit.2} as extracted from \cite{Cordier,Mane}. We use this information for guidance in estimating 
the slope $W'$ of the schematic ramp function that connects the $K\overline{K}$ spectrum with the onset of the QCD continuum in the range 
between $\sqrt{s} \simeq 1.4$ and $1.6$ GeV. 
The resulting interpolated full spectral function is also displayed in Fig.\,\ref{fig:vac.fit.2}. 
The schematic ramp function is parametrized as: 
\begin{equation}
\begin{split}
s_0 &= \frac{s_1 + s_2}{2}, \\
W' &= \frac{1}{s_2 - s_1}. 
\end{split}
\end{equation}
Guided by the experimental data shown in Fig. \ref{fig:vac.fit.2}, we find: 
$\sqrt{s_0} = 1.66 \pm 0.03\,\mathrm{GeV}$ and $W' = 0.8 \pm 0.2\,\mathrm{GeV}^{-2}$. 

Using Eq.(\ref{eq:ansatz.ramp}) instead of Eq.(\ref{eq:ansatz}), the finite energy sum rules are 
modified as 
\begin{align}
\int_0^{s_2} ds \,R_{\phi}(s) &= c_0\, s_0 + c_2, \label{eq:sm1.ramp} \\
\int_0^{s_2} ds\, s \,R_{\phi}(s) &= \frac{c_0}{2} \Big(s_0^2 + \frac{1}{12\, W'\,^2} \Big) - c_4, \label{eq:sm2.ramp} \\
\int_0^{s_2} ds \,s^2 \,R_{\phi}(s) &= \frac{c_0\,s_0}{3} \Big(s_0^2 + \frac{1}{4\, W'\,^2} \Big) + c_6, \label{eq:sm3.ramp}
\end{align}
with $s_2 = s_0 + 1/(2\,W')$. It is evident that the step function (\ref{eq:ansatz}) and the corresponding sum rules of Eqs.(\ref{eq:sm1}-\ref{eq:sm3}) are recovered in the limit $W' \to \infty$. 
\begin{figure*}
\begin{center}
\includegraphics[width=7cm]{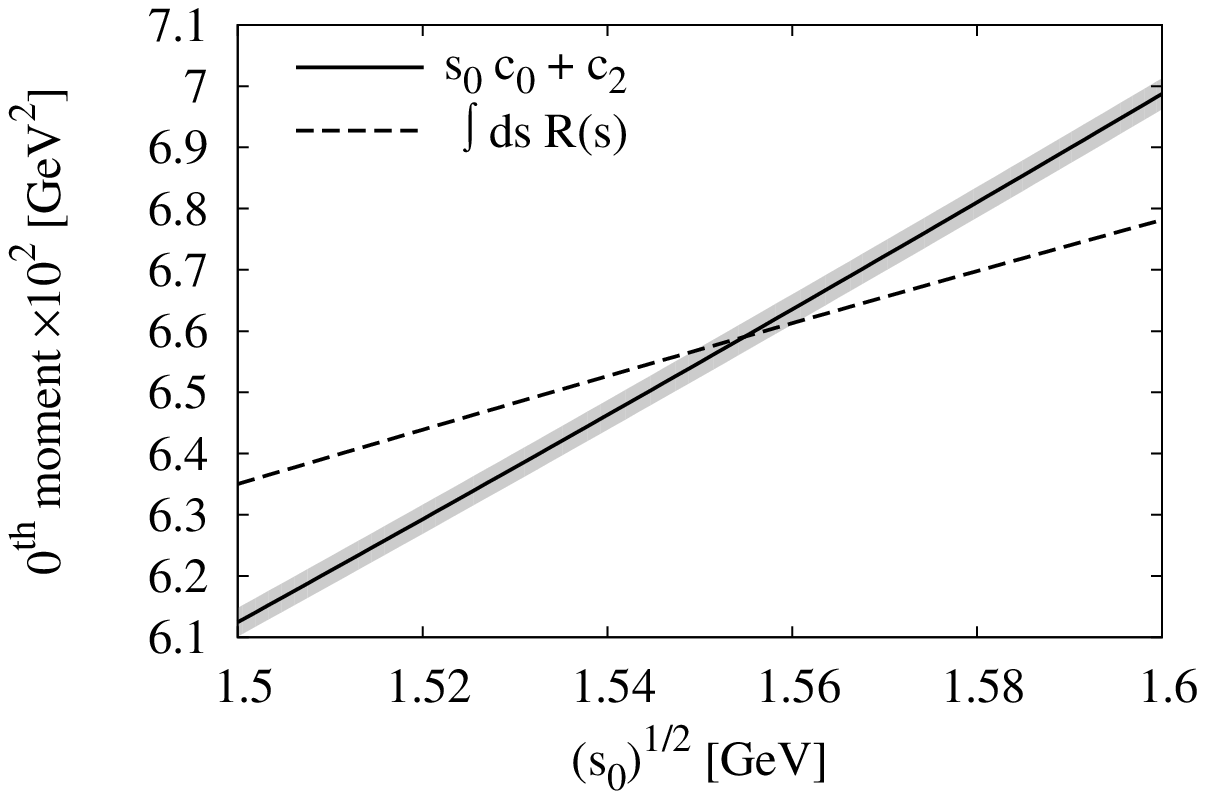}
\includegraphics[width=7cm]{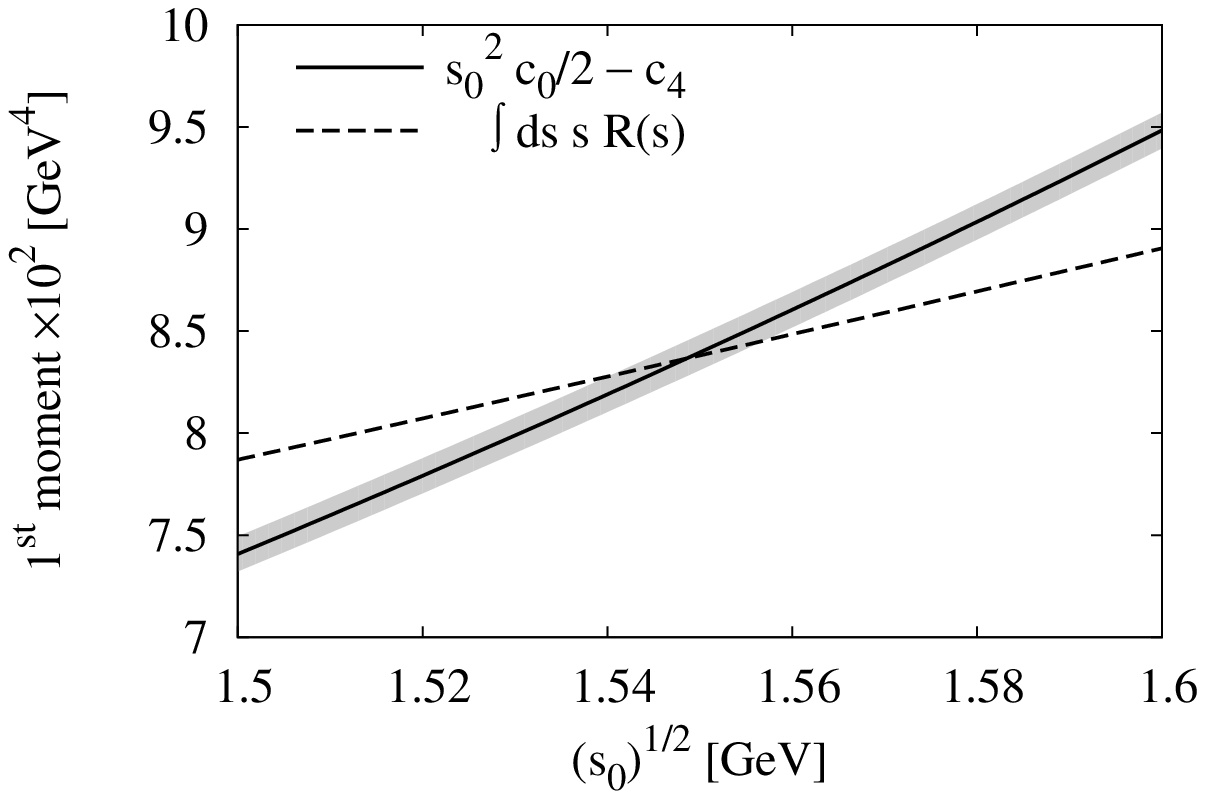}
\end{center}
\vspace{-0.6cm}
\caption{The left-hand and right-hand sides of Eqs.\,(\ref{eq:sm1}) and (\ref{eq:sm2}) as functions of the delineation scale $\sqrt{s_0}$. 
The error bands (printed in gray) are extracted from the uncertainties of the parameters given in Table \,\ref{tab:parameters}.}
\label{fig:vac.moments}
\end{figure*}

\subsection{\label{Momentanalysisvac} Matching spectral function and sum rules}
In the next step the moments of the spectral function (\ref{eq:veccorr3}) are analyzed, first using the generic sum rules (\ref{eq:sm1},\ref{eq:sm2}) 
with step-function onset of the perturbative QCD continuum. 
Here we follow Refs.\,\cite{Marco,Kwon} and substitute Eq.\,(\ref{eq:veccorr3}) into the left-hand sides of Eqs.\,(\ref{eq:sm1}) 
and (\ref{eq:sm2}). The ensuing equations are then solved individually for $s_0$. 

In Fig.\,\ref{fig:vac.moments} the left- and right-hand sides of Eqs.\,(\ref{eq:sm1}) and (\ref{eq:sm2}) are compared. 
The  parameter values appearing on the right-hand sides are given in Table \ref{tab:parameters}. 
\begin{table}
\renewcommand{\arraystretch}{1.5}
\setlength{\tabcolsep}{10pt}
\begin{center}
\caption{Parameter values and their uncertainties used for the vacuum QCD sum rule analysis.} 
\label{tab:parameters}
\begin{tabular}{lc}  
\hline 
$\alpha_s(2~\mathrm{GeV})$ & $0.31\pm0.01$ \cite{Olive} \\
$m_s\,(2~\mathrm{GeV})$ & $95\pm5\,\,\mathrm{MeV}$ \cite{Olive} \\
$\langle \overline{s} s \rangle\,(2~\mathrm{GeV})$ & $(-290\pm15\,\,\mathrm{MeV})^3$ \cite{McNeile} \\
$\big \langle \frac{\alpha_s}{\pi} G^2 \big \rangle$ & $0.012 \pm 0.004\,\,\mathrm{GeV}^4$ \cite{Colangelo} \\
\hline
\end{tabular}
\end{center}
\end{table} 
The renormalization scale for these parameters is taken at 2\,GeV, at which point the mass of the strange quark (and the strange quark condensate) 
are reasonably well determined according to Refs.\,\cite{Olive,McNeile}. This choice of scale is consistent with the 
energy at which the spectral function has reached the perturbative QCD limit (see Fig.\,\ref{fig:vac.fit.2}). The (perturbative) renormalization scale should be above 
the delineation scale $\sqrt{s_0}$, but otherwise its detailed value is not important. 
Solving the zeroth moment equation, we find $\sqrt{s_0} = 1.55\pm0.01~\mathrm{GeV}$. 
For the first moment the resulting matching scale turns out to be exactly the same, 
$\sqrt{s_0} = 1.55\pm0.01~\mathrm{GeV}$. We thus observe a remarkable degree
of consistency between these two moments with respect to the delineation scale $s_0$ at which perturbative
QCD takes over. 

A quantity reflecting the $\phi$ mass together with some of the emerging continuum above the resonance can be 
defined by taking the ratio of the first and zeroth moments, each integrated up to a scale $\overline{s}$  with $m_\phi^2 < \overline{s} < s_0$. 
This ratio represents a squared mass averaged over the spectrum $R_{\phi}(s \leq \overline{s})$. 
Choosing $\sqrt{\overline{s}} = 1.2$ GeV turns out to be convenient as it lies on one hand well above the resonance peak, 
but on the other hand low enough for the average not to be too much diluted by the continuum.
Furthermore, the vacuum and in-medium spectral functions 
$R_\phi(s)$ become identical (and constant) above that scale as will be shown. With this $\overline{s} = (1.2$ GeV$)^2$ one finds
\begin{equation}
\overline{m}_{\phi} = \sqrt{\frac{\int_0^{\overline{s}}ds \, s \, R_{\phi}(s)}{\int_0^{\overline{s}}ds \, R_{\phi}(s)}} 
 \simeq 1038 ~\mathrm{MeV}.  
 \label{eq:ratio.vacuum}
\end{equation}
The presence of the continuum above the $\phi$ meson resonance in 
Fig.\,\ref{fig:vac.fit} implies that $\overline{m}_{\phi}$ 
is slightly larger than the physical $\phi$ mass, $m_{\phi}=1019\,\mathrm{MeV}$. 
The value of $\sqrt{\overline{s}}$ is of course just a matter of choice and the number given in 
Eq.(\ref{eq:ratio.vacuum}) depends on this choice. Our emphasis is not on the precise value 
of Eq.(\ref{eq:ratio.vacuum}), but rather on its modification when going from vacuum to nuclear matter. 

\begin{figure*}
\begin{center}
\includegraphics[width=8cm]{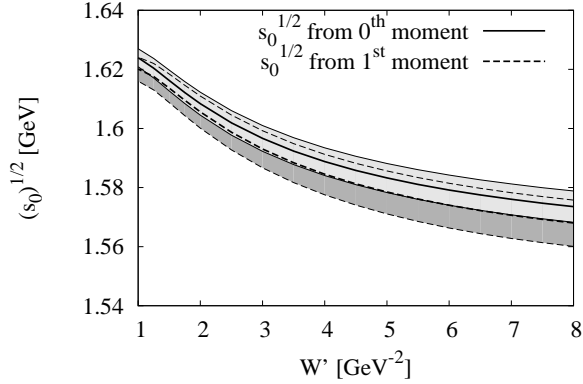}
\end{center}
\vspace{-0.6cm}
\caption{Solutions of Eqs.\,(\ref{eq:sm1.ramp},\ref{eq:sm2.ramp}), shown as a function of the ramp slope $W'$.}
\label{fig:ramp.1}
\end{figure*}

In order to study the dependence of our findings on the specific modeling of the continuum onset, we now replace the step function 
of Eq.\,(\ref{eq:ansatz}) by the schematic ramp function described in Eqs.\,(\ref{eq:ansatz.ramp}) and (\ref{eq:ansatz.ramp2}). 
The corresponding sum rules of Eqs.\,(\ref{eq:sm1.ramp}) and (\ref{eq:sm2.ramp}) 
are then solved for $s_0(W')$, keeping the slope of the ramp ($W'$) as a free parameter. 
Results are summarized in Fig.\,\ref{fig:ramp.1}. 
As seen in this figure the solutions $s_0(W')$ both for the zeroth and first moment increase moderately with 
decreasing ramp slope $W'$ but remain consistent within uncertainties unless this slope is too soft. 
This behavior demonstrates that the detailed modeling of the continuum onset has only a weak impact on the moment analysis 
and (for $W' \gtrsim 1\,\mathrm{GeV}^{-2}$) does not diminish the consistency between the sum rules for zeroth and first moments.   

\section{\label{Nuclmatter} Spectral moment analysis in nuclear matter}
\subsection{Spectral function at finite density} 
Next we extend the FESR analysis to the $\phi$ meson in nuclear matter. 
To leading (linear) order in the density $\rho$, the in-medium $\phi$ meson self-energy can be 
written: 
\begin{equation}
\Pi_{\phi}(q_0,\,\bm{q};\rho) = \Pi_{\phi}^{\mathrm{vac}}(q^2) - \rho \mathcal{T}_{\phi N} (q_0,\,\bm{q}), 
\label{eq:phi.nucl}
\end{equation}
with $q^2 = q_0^2 - \bm{q}^2$, recalling that $q_0$ and $\bm{q}$ can be varied independently in the nuclear medium. Here $\mathcal{T}_{\phi N}$ is the free forward $\phi$-nucleon scattering amplitude. It involves the $\phi\rightarrow K\overline{K}$ transition followed by the interaction of the kaon (antikaon) with a nucleon \cite{Klingl3}. 
In the present work we stay in the low-density limit, i.e. at leading order in $\rho$, for reasons of consistency with the sum rule analysis
which is also restricted to in-medium quark and gluon condensates linear in the density. Studies beyond linear order in $\rho$ have been performed in Refs. \cite{Cabrera,LutzKorpa,Cieply}.

We focus now first on a $\phi$ at rest in the medium $(\bm{q} = 0)$ and study S-wave $KN$ and $\overline{K}N$ interactions in $\mathcal{T}_{\phi N} (q_0,\,\bm{q}=0)$. 
To compute the respective $\phi$-nucleon scattering amplitude, consider the 
kaon and antikaon propagators in nuclear matter. For the $K^+K^-$ case they are written in spectral representations as 
 \begin{equation}
D_{K^{\pm}}(p_0,\bm{p};\rho) = \int_{0}^{\infty} du_0 
\left( \frac{A^{\pm}(u_0,\bm{p})}{p_0 -u_0 +i\epsilon} - \frac{A^{\mp}(u_0,\bm{p})}{p_0 +u_0 -i\epsilon} \right), 
\label{eq:eq1}
\end{equation}
with 
\begin{equation}
A^{\pm}(p_0,\bm{p}) = -\frac{1}{\pi} \frac{\mathrm{Im}\, \Sigma_{K^{\pm}}(p_0,\bm{p};\rho)}
{\left[p_0^2 -\bm{p}^2 - m_{K^{\pm}} -\mathrm{Re}\,\Sigma_{K^{\pm}}(p_0,\bm{p};\rho) \right]^2
+ \left[\mathrm{Im}\,\Sigma_{K^{\pm}}(p_0,\bm{p};\rho) \right]^2 }.  
\label{eq:eq2}
\end{equation}
Here $\Sigma_{K^{\pm}}(q_0,\bm{q};\rho)$ denote the in-medium  $K^+$ and $K^-$ self-energies. For the 
$\phi\rightarrow K_0 \overline{K}_0$ case the corresponding $K^0$ and $\overline{K}^0$ propagators have an analogous structure, with $A^+$ replaced by $A^0$ and so forth. 
The in-medium kaon self-energies involve the forward $KN$ and 
$\overline{K}N$ scattering amplitudes. For $KN$ (i.e. $K^+ N$ or $K^0 N$), the interaction is known to be repulsive and relatively weak. 
The corresponding amplitudes have small imaginary parts and depend weakly on energy in the relevant region around the 
$\phi$ meson pole \cite{Waas,Lutz1}. We can therefore approximate these amplitudes as real constants. 
The corresponding spectral functions $A^{+}$ and $A^{0}$ are simple $\delta$-functions in this approximation: 
\begin{equation}
A^{+}(p_0,\bm{p};\rho) = \delta \big(p^2 - m^{\star \,2}_{K^{+}}(\rho) \big),  
\label{eq:Aplus}
\end{equation}
and
\begin{equation}
A^{0}(p_0,\bm{p};\rho) = \delta \big(p^2 - m^{\star \,2}_{K^{0}}(\rho) \big), 
\label{eq:A0}
\end{equation}
with in-medium kaon masses $m^{\star}_K(\rho)$.
Following \cite{Klingl3} and taking into account the repulsive nature of the $KN$ interaction, we use  
$m^{\star}_{K}(\rho) \simeq m_K + \Delta m_K(\rho/\rho_0)$ with $m_K = 496$ MeV, the average of the $K^+$ and $K^0$ masses, and $\Delta m_K \simeq 39$ MeV. 
This gives $m^{\star}_K(\rho = \rho_0) \simeq 535\,\mathrm{MeV}$ for both $K^{+}$ and $K^{0}$ at normal nuclear matter density $\rho_0$, and we 
assume $m^{\star}_K$ to be a linear function of the density $\rho$. 

For the $\overline{K}N$ ($K^- N$ or $\overline{K}^0 N$) system, the situation is more involved as the 
attractive $\overline{K}N$ interaction leads to the dynamical generation of the $\Lambda(1405)$, inducing 
a strong energy-dependence and a large imaginary part onto the scattering amplitudes. To take these features 
properly into account, we employ amplitudes derived from $SU(3)$ chiral effective field
theory \cite{Ikeda,Ikeda2}, including $\overline{K}N\leftrightarrow\pi\Sigma$ coupled channels and the $\Lambda(1405)$ resonance. These amplitudes reproduce all presently available scattering data together with accurate kaonic hydrogen measurements. 
In terms of the antikaon self energies, we have for symmetric nuclear matter: 
\begin{equation}
\Sigma_{K^{-}}(p_0,\bm{p};\rho) \simeq - 2 \pi \rho \frac{\sqrt{s}}{M _N} \Big[ f(K^{-} p \to K^{-} p, \sqrt{s}) 
+ f(K^{-} n \to K^{-} n, \sqrt{s}) \Big],
\end{equation} 
and 
\begin{equation}
\Sigma_{\overline{K}^{0}}(p_0,\bm{p};\rho) \simeq - 2 \pi \rho \frac{\sqrt{s}}{M _N}\Big[f(\overline{K}^{0} p \to \overline{K}^{0} p, \sqrt{s}) 
+ f(\overline{K}^{0} n \to \overline{K}^{0} n, \sqrt{s}) \Big].
\end{equation}
Here $\sqrt{s}$ stands for the kaon-nucleon energy. In the lab frame with a nucleon at rest it is given as 
\begin{equation}
\sqrt{s} = \sqrt{M_N^2 + m_K^2 + 2M_N\, p_0}.
\end{equation}
where $M_N$ ist the nucleon mass and we neglect the small proton-neutron mass difference.

With the kaon and antikaon propagators just outlined, the next step is to evaluate the kaon loops in nuclear matter as they contribute to the $\phi$ meson self-energy. Considering the $K^+K^-$ loop for a $\phi$ at rest and introducing again the notation $q_0 \equiv \omega$ for convenience, one obtains
\begin{equation}
\Pi_{\phi \to K^+ K^-}(\omega,\,\bm{q}=0;\rho) = \frac{2ig^2}{3} \int \frac{d^4 p }{(2\pi)^4 }\, \,\bm{p}^2\,\, 
D_{K^{+}}(p_0,\bm{p};\rho) \,\,D_{K^{-}}(\omega-p_0,-\bm{p};\rho).
\end{equation}
We have omitted a tadpole contribution that is not relevant here. Substituting Eq.\,(\ref{eq:eq1}) 
into the integrand, the imaginary part of $\Pi_{\phi \to K^+ K^-}$ can be evaluated as 
\begin{equation}
\begin{split}
&\mathrm{Im}\,\Pi_{\phi \to K^+ K^-}(\omega,\,\bm{q}=0;\rho) \\
=& - \frac{g^2}{96 \pi\,\omega^4} 
\int_{-\bm{p}^2}^{\infty} du_{+}^2  \int_{-\bm{p}^2}^{\infty} du_{-}^2\,\, 
A^{+}\Big(\sqrt{u^2_{+} + \bm{p}^2}, \bm{p}; \rho \Big)\,\, A^{-}\Big(\sqrt{u^2_{-} + \bm{p}^2}, \bm{p}; \rho \Big)\, 
\lambda^{3/2}\big(\omega^2, u^2_{+}, u^2_{-}\big), 
\end{split}
\end{equation}
where $\bm{p}^2 = \lambda \big(\omega^2, u^2_{+}, u^2_{-}\big) / (4 \omega^2)$ and 
$\lambda(a, b, c) = a^2 + b^2 + c^2 - 2ab - 2ac - 2bc$ is the K\"allen function. Making use of Eq.\,(\ref{eq:Aplus}), this 
can be further simplified to 
\begin{equation}
\mathrm{Im}\,\Pi_{\phi \to K^+ K^-}(\omega,\,\bm{q}=0;\rho) 
= - \frac{g^2}{96 \pi\,\omega^4} 
\int_{-\bm{p}^2}^{\infty} du_{-}^2 \,\, A^{-}\Big(\sqrt{u^2_{-} + \bm{p}^2}, \bm{p}; \rho \Big)\,\, 
\lambda^{3/2}\big(\omega^2, m^{\star \,2}_{K^{+}}(\rho), u^2_{-}\big) 
\end{equation}
and $\bm{p}^2 = \lambda \big(\omega^2, m^{\star \,2}_{K^{+}}(\rho), u^2_{-}\big) / (4 \omega^2)$. 
The final integration over $u_{-}^2$ is then performed numerically. As before, analogous expressions can be 
derived for the $\overline{K}^0 K^0$ loop. 

The above expression of $\mathrm{Im}\,\Pi_{\phi \to K^+ K^-}(\omega,\,\bm{q}=0;\rho)$ includes, by construction, finite density contributions beyond leading linear order in $\rho$. For consistency, the imaginary part of the (free) forward $\phi$-nucleon scattering amplitude in Eq.\,(\ref{eq:phi.nucl}) is then constructed by taking the zero-density limit:  
\begin{equation}
\mathrm{Im}\,\mathcal{T}_{\phi N}(\omega,\,\bm{q}=0) =  \lim_{\rho \to 0} \frac{\mathrm{Im}\,\Pi^{\mathrm{vac}}_{\phi}(\omega,\,\bm{q}=0) 
- \mathrm{Im}\,\Pi_{\phi}(\omega,\,\bm{q}=0;\rho)}{\rho}. 
\label{eq:eq7}
\end{equation}
Finally, the real part of $\mathcal{T}_{\phi N}(\omega,\,\bm{q}=0)$ is computed using
a once subtracted dispersion relation: 
\begin{equation}
\mathrm{Re}\,\mathcal{T}_{\phi N}(\omega,\,\bm{q}=0) = c + \frac{\omega^2}{\pi} \mathcal{P} 
\int_0^{\infty} du^2 \,\,\frac{\mathrm{Im}\,\mathcal{T}_{\phi N}(u,\,\bm{q}=0)}
{u^2(u^2 - \omega^2)}, 
\label{eq:eq9}
\end{equation}
in which the subtraction constant is $c = 0$ according to the Thomson limit of Compton scattering at $\omega=0$. 
Given both real and imaginary parts of $\mathcal{T}_{\phi N}$ we can now return to Eq.\,(\ref{eq:phi.nucl}) 
for the $\phi$ meson self-energy and finally to Eq.\,(\ref{eq:veccorr3}) for computing the spectral function. 

P-wave $KN$ and $\overline{K}N$ interactions are incorporated using the spectral function provided in \cite{Klingl3}. 
This spectral function includes the relevant baryon octet and decuplet intermediate states at 
one-loop level. It describes the region around the $\phi$ meson peak up to $\omega \simeq 1.1$ GeV and is then smoothly 
connected to the continuum at higher energies. The $\phi$ spectrum involving pure S-wave $KN$ and $\overline{K}N$ couplings 
starts at the $K\overline{K}$ threshold, located at twice the kaon mass in vacuum and shifted downward in the medium primarily by 
the attractive $\overline{K}N$ interaction. The inclusion of P-wave interactions couples the $\phi$ to the $YK$ continuum with correspondingly 
lower thresholds, where $Y$ stands for $\Lambda, \Sigma$ and $\Sigma^*$.
 
\begin{figure}
\begin{center}
\includegraphics[width=10cm]{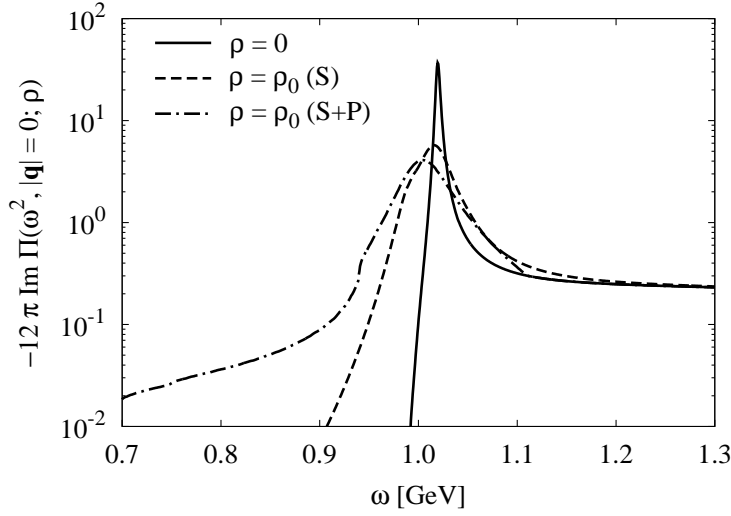}
\end{center}
\vspace{-0.6cm}
\caption{The spectral function $-12 \pi \mathrm{Im} \Pi(\omega^2)$ in vacuum (solid curve) and at normal nuclear matter density, $\rho = \rho_0 = 0.17$ fm$^{-3}$ 
(only S-waves: dashed curve, S- and P-waves: dash-dotted curve).}
\label{fig:vac}
\end{figure}

The vacuum and in-medium spectral functions computed in this way are plotted as solid (vacuum), dashed (only S-wave) and 
dash-dotted (S- and P-wave) curves in Fig.\,\ref{fig:vac}. 

\subsection{The operator product expansion at finite density}
Finite density effects modify the OPE input for the sum rules. 
At leading order in the density $\rho$, 
the vacuum condensates receive the following corrections: 
\begin{align}
\langle \overline{s} s \rangle_{\rho} \simeq &\; \langle \overline{s} s \rangle + \langle N| \overline{s} s |N \rangle \rho = \langle \overline{s} s \rangle + \frac{\sigma_{sN}}{m_s} \, \rho, \\
\Big \langle \frac{\alpha_s}{\pi} G^2 \Big \rangle_{\rho} \simeq &\; \Big \langle \frac{\alpha_s}{\pi} G^2 \Big \rangle 
+ \Big \langle N \Big| \frac{\alpha_s}{\pi} G^2 \Big|N\Big \rangle \rho 
= \; \Big \langle \frac{\alpha_s}{\pi} G^2 \Big \rangle - \frac{8}{9}\big(M_N - \sigma_{\pi N} - \sigma_{sN}\big)\,\rho. 
\end{align} 
Here $M_N$ is the physical nucleon mass, $\sigma_{\pi N} = 2m_q \langle N| \overline{q} q |N \rangle$ stands for the $\pi N$ sigma term and 
$\sigma_{sN} = m_s \langle N| \overline{s} s |N \rangle$ for the strangeness sigma term of the nucleon. 
There is furthermore a correction coming from a twist-2 operator: 
\begin{align}
\mathcal{S}\mathcal{T} \langle N| \bar{s}\, \gamma^{\,\mu} D^{\nu} s |N \rangle = \frac{-i}{2M_N} A^s_2\, \big(p^{\mu} p^{\nu} - \frac{1}{4} M_N^2 \,g^{\mu \nu} \big). 
\end{align} 
The symbols $\mathcal{S}\mathcal{T}$ make the matrix symmetric and traceless with respect to the Lorentz indices, and $p^{\mu}$ is the four-momentum of the nucleon ($p^2 = M_N^2$). The quantity
$A^s_2$ is the first moment of the strange quark distributions in the nucleon\footnote{An additional term related to the first moment 
of the gluon distribution \cite{Gubler} is ignored here for simplicity. It adds to the uncertainty of the analysis and will be discussed later}
\begin{align}
A^s_2 = 2 \int_0^1 dx~x \big[s(x) + \overline{s}(x) \big]. 
\end{align} 
The resulting in-medium corrections to the coefficient $c_4$ of Eq.\,(\ref{eq:c4}) read  
\begin{align}
\delta c_4  = \left[\left(A^s_2 - \frac{2}{27}\right) M_N + \frac{2}{27}\left(28 \sigma_{sN} + \sigma_{\pi N}\right) \right] \rho, 
\label{eq:finite.density.c4}
\end{align}
while $c_0$ and $c_2$ respectively remain at their vacuum value. 
Our choices for the parameter values needed to evaluate Eq.(\ref{eq:finite.density.c4}) are given in Table \ref{tab:parameters.matter}. 
\begin{savenotes}
\begin{table}
\renewcommand{\arraystretch}{1.5}
\setlength{\tabcolsep}{10pt}
\begin{center}
\caption{Parameter values and ranges used as input for the QCD sum rule analysis in nuclear matter. The moment $A^s_2$ of the strange quark parton distribution is given at a renormalization scale $\mu = 1$ GeV. 
(The $25\,\%$ error in $A^s_2$ covers possible uncertainties related 
to an evolution towards $\mu = 2\, \mathrm{GeV}$, the renormalization scale at which the parameters in Table \ref{tab:parameters} 
are determined.)} 
\label{tab:parameters.matter}
\begin{tabular}{lc}  
\hline 
$M_N$ & $940\,\,\mathrm{MeV}$ \\
$\sigma_{sN}$ & $35\pm35\,\,\mathrm{MeV}$ 
\cite{Young,Babich,Durr,Horsley,Bali,Semke,Freeman,Shanahan,Ohki,Alarcon1,Engelhardt,Junnarkar,Jung,Gong,Alexandrou,Lutz,Ren,
DurrBMW,YangchiQCD,AbdelRehimETM}\footnote{Note however that one of the recent lattice QCD results \cite{DurrBMW} points to a larger 
value for $\sigma_{sN}$. See the discussion in Section \ref{Sigma}.} \\
$\sigma_{\pi N}$ & $45\pm7\,\,\mathrm{MeV}$ \cite{Gasser}\footnote{According to recent analysis \cite{Hoferichter,Hoferichter2,Alarcon}, $\sigma_{\pi N}$ may increase to a somewhat larger value, 
but could also be smaller \cite{DurrBMW} (see also the recent discussion in \cite{Hoferichter3}). Whatever the true value turns out to be, it will not affect the conclusions of this work.} \\
$A^s_2$ & $0.044 \pm 0.011$ \cite{Martin,Gubler} \\
\hline
\end{tabular}
\end{center}
\end{table}
\end{savenotes}
For the strangeness sigma term, $\sigma_{sN}$, no generally accepted value is available at present and we hence take a rather broad  range that is consistent with most of the recent studies \cite{Young,Babich,Durr,Horsley,Bali,Semke,Freeman,Shanahan,Ohki,Alarcon1,Engelhardt,Junnarkar,Jung,Gong,Alexandrou,Lutz,Ren,DurrBMW,YangchiQCD,AbdelRehimETM}, 
which include direct lattice QCD computations and chiral extrapolations of available lattice data. 

Note that the terms proportional to the nucleon mass enter with opposite signs and partly cancel. A special feature of the $\phi$ meson in-medium sum rule is the strong weight (by the factor $28$) on the strangeness sigma term $\sigma_{sN}$ relative to $\sigma_{\pi N}$. The first spectral moment of the $\phi$ in nuclear matter is linearly connected to $\sigma_{sN}$ and hence to the scalar density of strange quarks in the nucleon. This pronounced sensitivity of $\delta c_4$ with respect to $\sigma_{sN}$ is of particular interest and will be discussed separately in Section \ref{Sigma}. 

\subsection{Moment analysis with a step-function continuum onset}
\begin{figure*}
\begin{center}
\includegraphics[width=7cm]{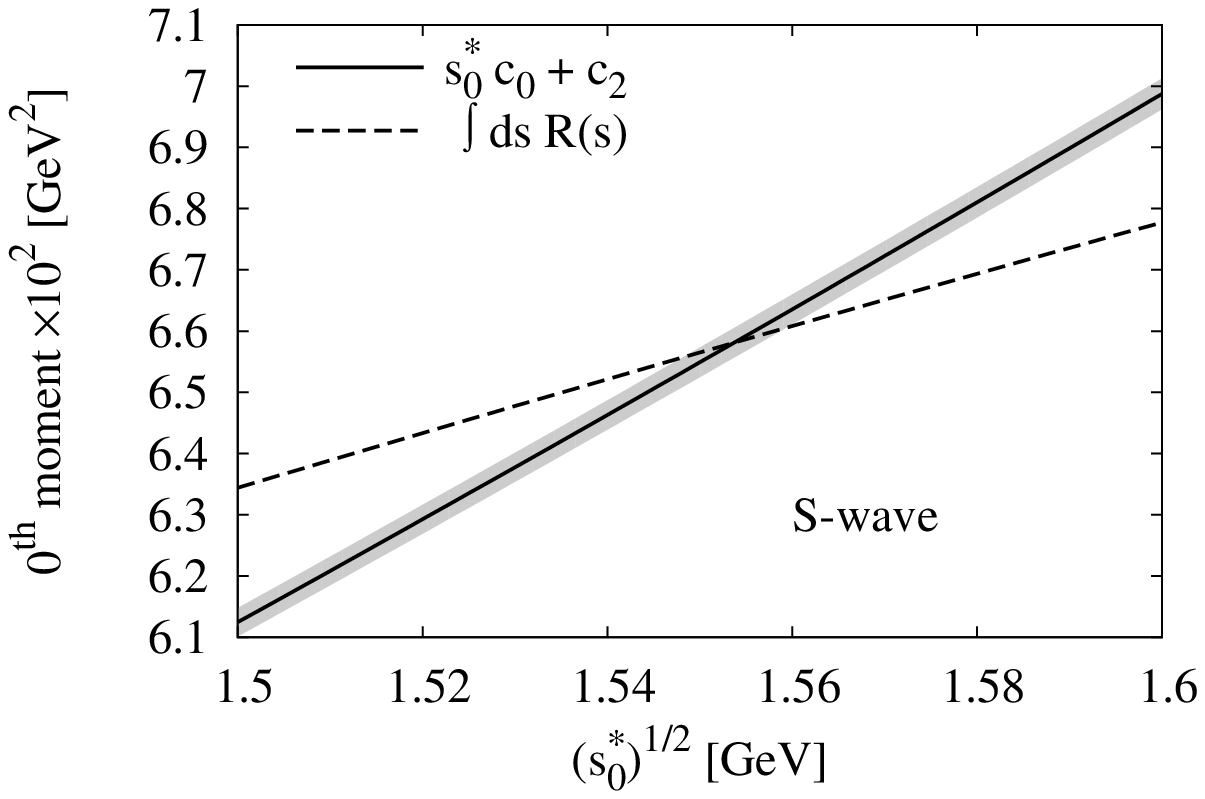}
\includegraphics[width=7cm]{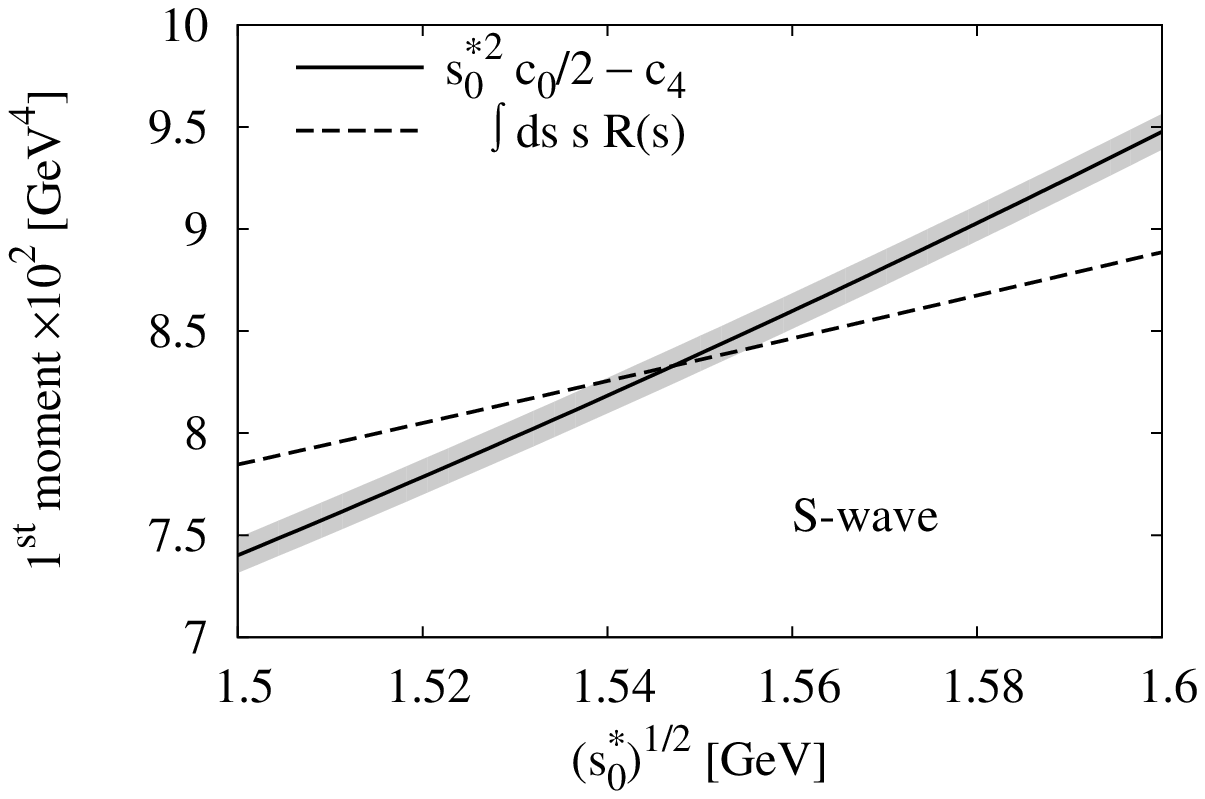}
\includegraphics[width=7cm]{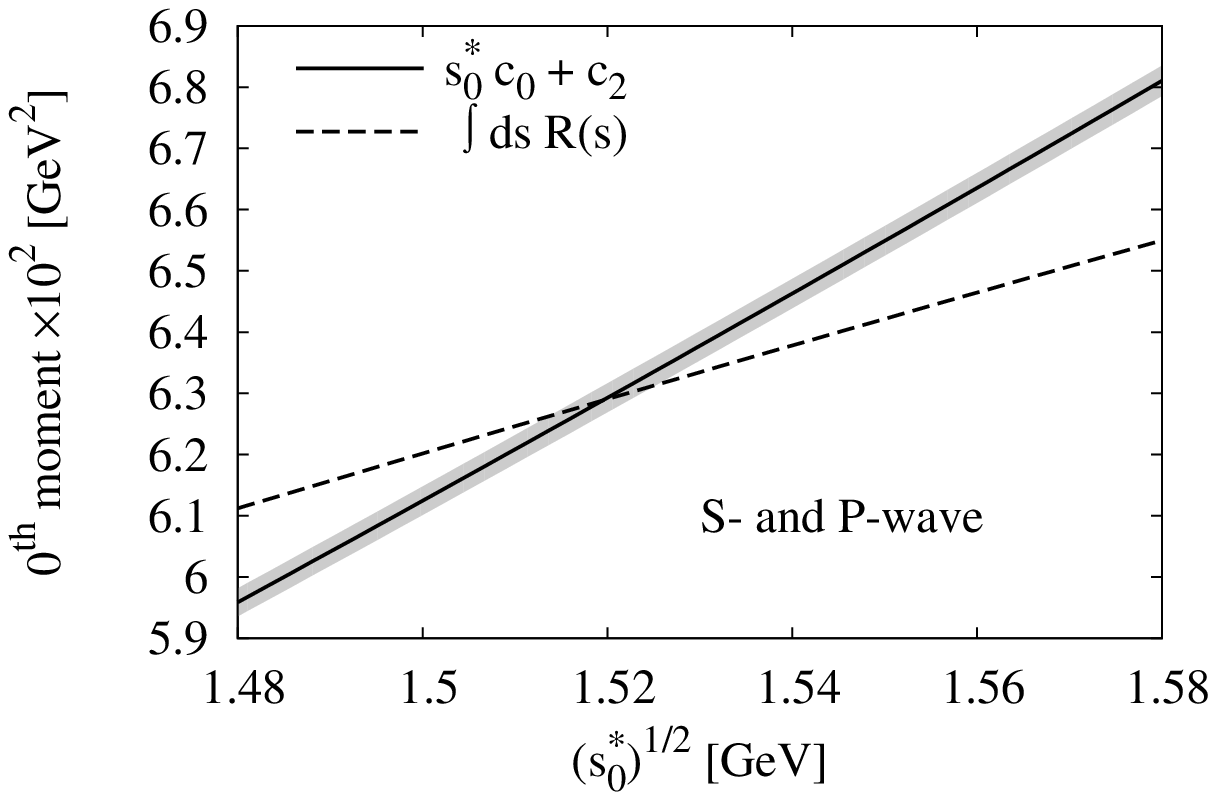}
\includegraphics[width=7cm]{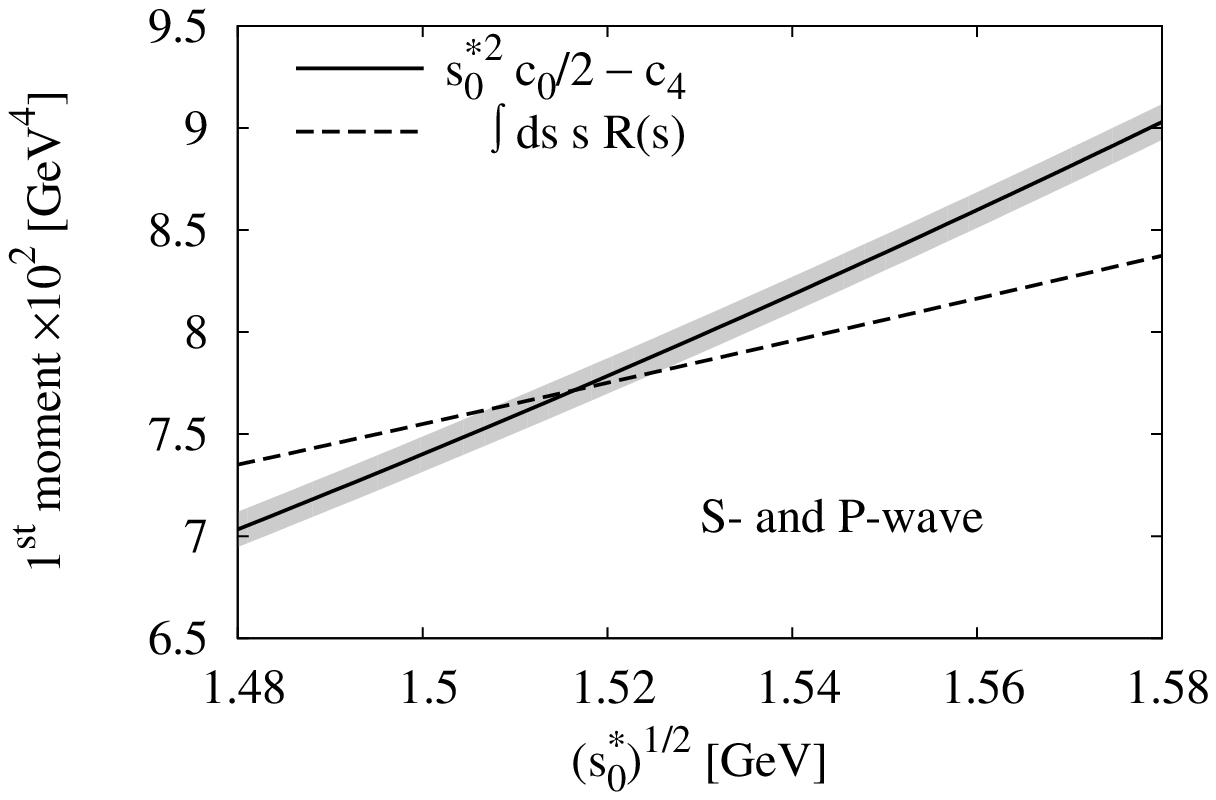}
\end{center}
\vspace{-0.6cm}
\caption{The left- and right-hand sides of the nuclear matter versions of Eqs.(\ref{eq:sm1}) and (\ref{eq:sm2}) as functions of $\sqrt{s_0^{\ast}}$ at normal nuclear matter density, $\rho = \rho_0 = 0.17$ fm$^{-3}$.}
\label{fig:matter.moments}
\end{figure*}

With the input quantities listed in Table \ref{tab:parameters}, the analysis of in-medium spectral moments is now carried out in the same way as in the previous section. 
As a first step, consider only S-wave kaon- and antikaon-nucleon interactions, using the dashed curve in Fig.\,\ref{fig:vac} as input.
We solve again Eqs.\,(\ref{eq:sm1}) and (\ref{eq:sm2}) for the in-medium delineation scale between low-energy and perturbative QCD regions, now denoted $s_0^{\ast}$ in order to distinguish it  
from the vacuum value $s_0$. The left- and right-hand sides of the equations are shown in the upper plots of Fig.\,\ref{fig:matter.moments} as functions of $\sqrt{s_0^{\ast}}$. 
At normal nuclear matter density, $\rho = \rho_0 = 0.17$ fm$^{-3}$, one finds
$\sqrt{s_0^{\ast}} = 1.55\pm0.01\,\mathrm{GeV}$ from the zeroth moment, and 
$\sqrt{s_0^{\ast}} = 1.55\pm0.01\,\mathrm{GeV}$ from the first moment. At this point both in-medium scales are again fully consistent with each other and show no change in comparison with the corresponding vacuum values. 

Setting once again $\overline{s} = (1.2\,\mathrm{GeV})^2$ for convenience as in the vacuum case, an in-medium mass measure averaged over the $\phi$ resonance and part of the continuum is given by the ratio 
\begin{equation}
\overline{m}_{\phi}^{\ast}(\rho=\rho_0) = \sqrt{\frac{\int_0^{\overline{s}}ds \,s \,R_{\phi}(s)}{\int_0^{\overline{s}}ds \,R_{\phi}(s)}} 
 \simeq 1035 ~\mathrm{MeV}.  
\end{equation}
Comparing this result with Eq.\,(\ref{eq:ratio.vacuum})
it is seen that the averaged mass $\overline{m}_{\phi}^{\ast}$ at $\rho = \rho_0$ does not deviate from the vacuum value apart from a marginal downward shift. 
In contrast, the resonant $\phi$ meson peak experiences a significant broadening, with a width of $\Gamma_{\phi} \simeq 24\,\mathrm{MeV}$ at normal nuclear matter density. 
The width is determined by the imaginary part of the self-energy at the resonance maximum. 

Effects of P-wave antikaon-nucleon interactions are studied employing the dash-dotted curve of Fig.\ref{fig:vac}. 
The numerical analysis is performed as in the previous paragraphs and with the same OPE input. 
For the scale parameter $s_0^{\ast}$ at $\rho = \rho_0$, we extract $\sqrt{s_0^{\ast}} = 1.52\pm0.01\,\mathrm{GeV}$ from 
the zeroth moment, and $\sqrt{s_0^{\ast}} = 1.52\pm0.01\,\mathrm{GeV}$ from the first moment. 
The behavior of the left- and right-hand sides of Eqs.\,(\ref{eq:sm1}) and (\ref{eq:sm2}) 
are shown in the lower plots of  Fig.\,\ref{fig:matter.moments}. 

Finally, setting $\overline{s} = (1.2\,\mathrm{GeV})^2$ as before, the ratio of first to zeroth moments at $\rho = \rho_0$ including both S- and P-wave interactions is computed as:  
\begin{equation}
\overline{m}_{\phi}^{\ast}(\rho=\rho_0) = \sqrt{\frac{\int_0^{\overline{s}}ds\, s \,R_{\phi}(s)}{\int_0^{\overline{s}}ds\,R_{\phi}(s)}} 
 \simeq 1022~\mathrm{MeV}.  
\end{equation}
Comparing this result again with the vacuum value $\overline{m}_{\phi}$ of Eq.\,(\ref{eq:ratio.vacuum}), the averaged mass now experiences a modest downward 
shift by about $16\,\mathrm{MeV}$, more than for the pure S-wave case. This difference is explained by both 
a small shift of the $\phi$ resonance peak in combination with the pronounced low-energy continuum in the spectral function, caused by the opening of kaon-hyperon channels 
in the presence of P-wave interactions (see Fig.\,\ref{fig:vac}). 
The broadening of the in-medium $\phi$ resonance is further increased significantly, reaching $\Gamma_{\phi}(\rho =\rho_0) \simeq 45\,\mathrm{MeV}$, 
an order of magnitude larger than the vacuum width ($\Gamma_{\phi}(0) = 4.3\,\mathrm{MeV}$). 

\begin{figure*}
\begin{center}
\includegraphics[width=7cm]{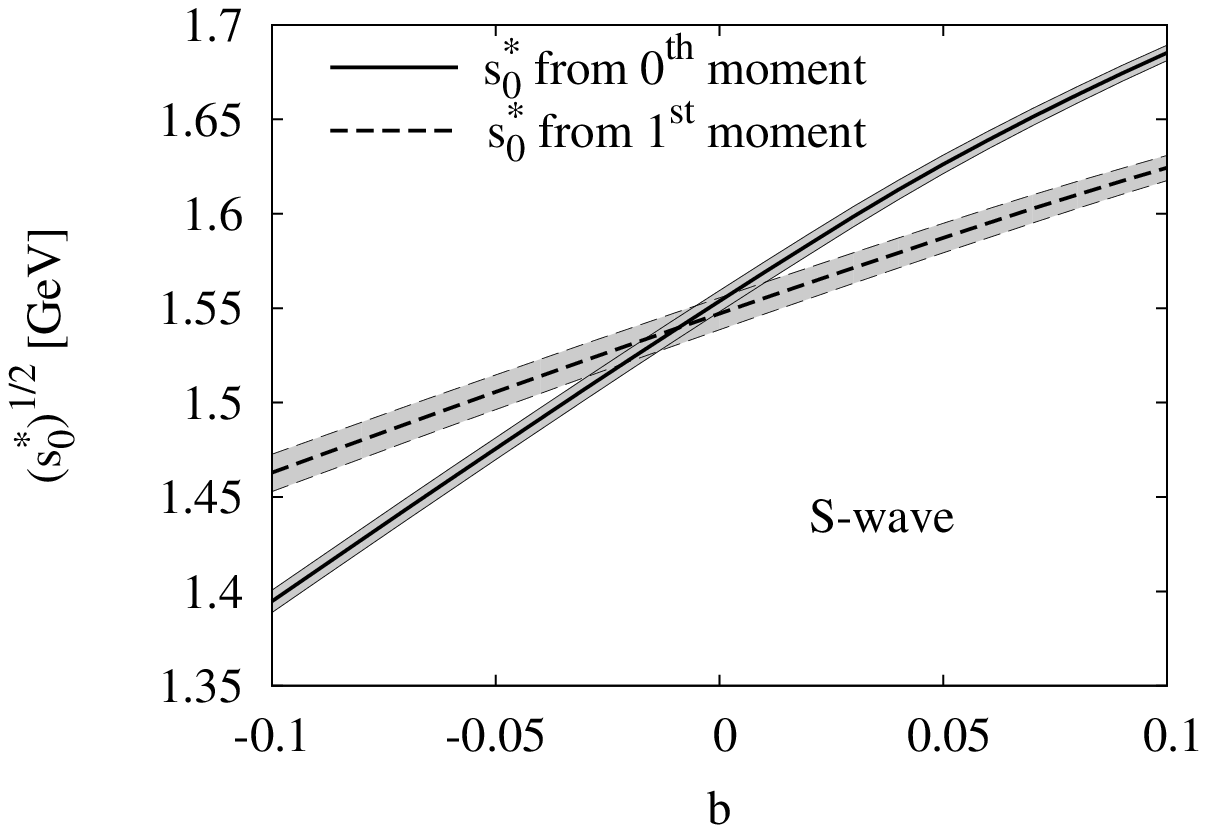}
\includegraphics[width=7cm]{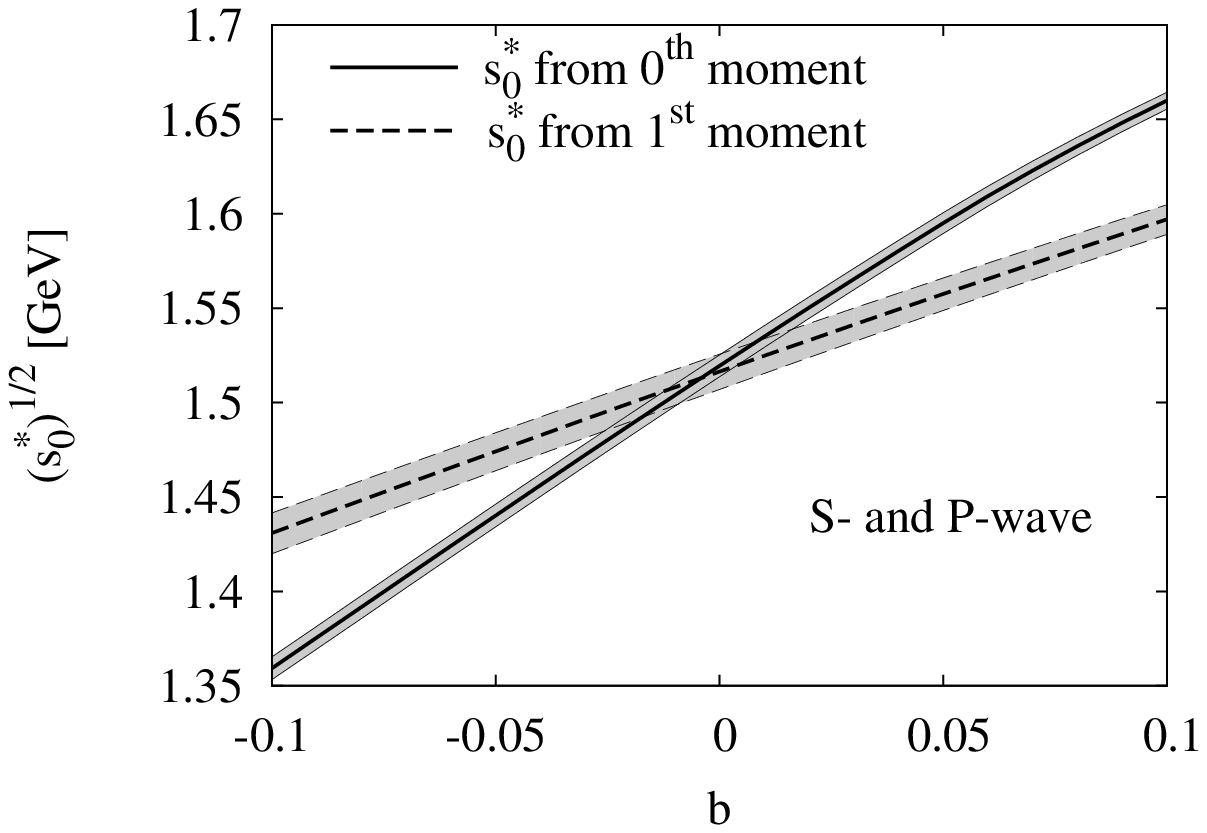}
\end{center}
\vspace{-0.6cm}
\caption{Solutions of zeroth and first moment sum rules of Eqs.\,(\ref{eq:sm1},\ref{eq:sm2}) at normal nuclear matter density, 
with a spectral function modified by an overall factor $1 - b \rho/\rho_0 = 1 - b$ according to Eq.\,(\ref{eq:veccorr.changed.coupl}).}
\label{fig:overall.factor}
\end{figure*}

The implicit assumption made in Eq.\,(\ref{eq:phi.nucl}) is that the form, Eq.\,(\ref{eq:veccorr3}), of the vacuum 
self-energy remains unchanged in the medium: a possible additional density dependence of the strange vector-current 
coupling to the $\phi$ meson channel is neglected. Here we briefly examine the validity of this assumption by 
introducing an extra factor, $1 - b \rho/\rho_0$, in Eq.\,(\ref{eq:veccorr3}), 
incorporating such a possible density dependence to leading order in $\rho$: 
\begin{equation}
\mathrm{Im}\Pi(q^2) = \Big( 1 - b \frac{\rho}{\rho_0} \Big)\frac{\mathrm{Im}\,\Pi_{\phi}(q^2)}{q^2 g_{\phi}^2} 
\Bigg| \frac{(1-a_{\phi})q^2 - \mathring{m}_{\phi}^2}{q^2 - \mathring{m}_{\phi}^2 - \Pi_{\phi}(q^2)} \Bigg|^2.
\label{eq:veccorr.changed.coupl}
\end{equation}
Performing the same sum rule analysis as before at normal nuclear matter density, it turns out that consistency of the 
spectral function with the FESR requires $b$ to be small. 
This is illustrated in Fig.\,\ref{fig:overall.factor}, in which the solutions $\sqrt{s_0^{\ast}}$ of the zeroth and first momentum sum rules 
are shown as a function of the parameter $b$, both for only S-wave and S+P-wave $\overline{K}N$ interaction cases. 
It is seen in these plots that for the two solutions to agree within their uncertainties, $b$ must lie in the range $|b| \lesssim 0.02$, 
which implies that 
the leading density dependence in Eq.\,(\ref{eq:phi.nucl}) is indeed well represented by the $\rho \mathcal{T}_{\phi N}$ 
term. 

\subsection{Analysis employing a ramp function}
\begin{figure*}
\begin{center}
\includegraphics[width=7cm]{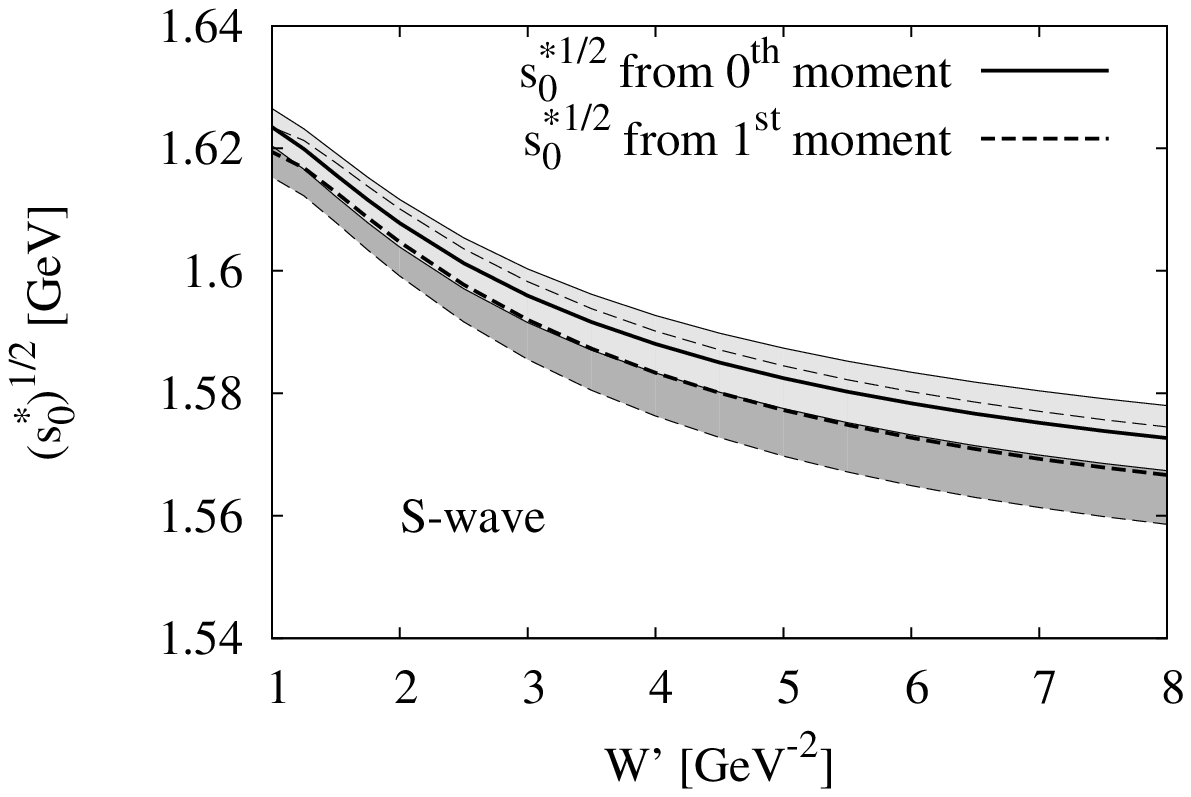}
\includegraphics[width=7cm]{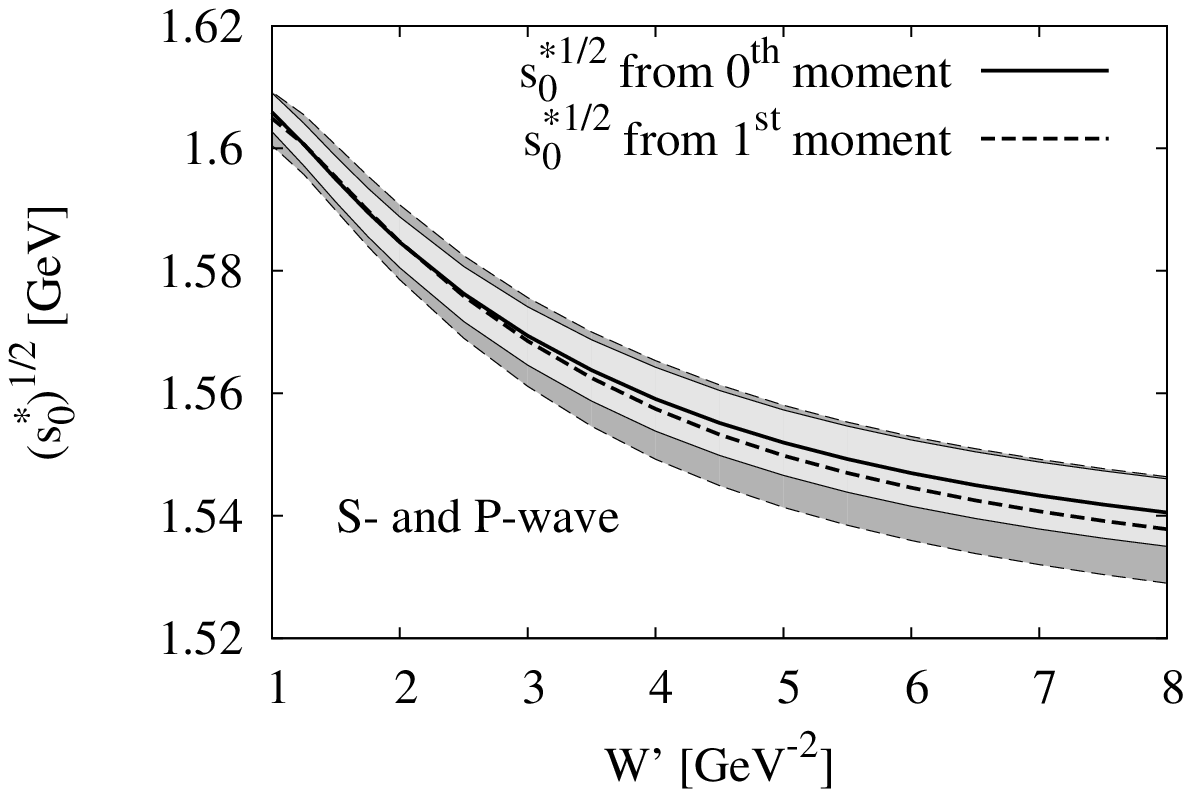}
\end{center}
\vspace{-0.6cm}
\caption{The solutions of Eqs.(\ref{eq:sm1.ramp}) and (\ref{eq:sm2.ramp}) at finite density, shown as a function of the slope parameter $W'$. In the 
left (right) plot, S-wave (S- and P-wave) $\overline{K}N$ interactions are taken into account.}
\label{fig:ramp.2}
\end{figure*}

We now repeat the in-medium FESR analysis with an improved modeling of the continuum onset. As before, we 
treat the slope of the ramp, $W'$, as a free parameter and solve the sum rules of Eqs. (\ref{eq:sm1.ramp}) and 
(\ref{eq:sm2.ramp}) for $s_0^{\ast}$ within a broad range of $W'$. 
The solutions are shown in Fig.\,\ref{fig:ramp.2}. 
For only S-wave $\overline{K}N$ interactions, the threshold parameter $s_0^{\ast}$ 
remains almost the same as in the vacuum (see Fig.\,\ref{fig:ramp.1}), while additional P-wave contributions lead to 
a small but notable decrease of $s_0^{\ast}$. The solutions from the zeroth and first moment sum rules 
turn out again to be consistent within uncertainties as long as $W'\gtrsim 1$ GeV$^{-2}$. 

\section{\label{Sigma} In-medium spectral moment and strangeness sigma term of the nucleon}

As pointed out previously, Eq.\,(\ref{eq:finite.density.c4}) establishes a linear relationship between the first moment of the in-medium $\phi$ meson spectral function and the strangeness sigma term of the nucleon. In simple terms: a possible shift of the $\phi$ mass in nuclear matter sets constraints on the scalar density of strange quarks in the nucleon. In practice, of course, the situation is more involved because of the substantial asymmetric in-medium broadening of the $\phi$ resonance and requires a more detailed assessment of the spectral distribution including the continuum.

Returning to Eq.\,(\ref{eq:finite.density.c4}) together with Eq.\,(\ref{eq:sm2}) we introduce
\begin{align}
\Delta(\rho) = \int_0^{s_0} ds\,s\,R_\phi(s;\rho) - {c_0\over 2}\,s_0^2 = -(c_4 + \delta c_4(\rho))
\label{eq:deltarho}
\end{align}
Subtracting the perturbative QCD limit, $c_0\,s_0^2/2$, from the finite integral over $sR_\phi(s)$ implies that the quantity $\Delta(\rho)$ is focused on the $\phi$ resonance region and its neighbourhood and does not depend on the scale $s_0$: both $c_4$ of Eq.\,(\ref{eq:c4}) and the density-dependent correction $\delta c_4$ of 
Eq.\,(\ref{eq:finite.density.c4}) are independent of $s_0$. Consequently, 
\begin{align}
\Delta(\rho=0) - \Delta(\rho)= \delta c_4(\rho) = \left[\left(A_2^s - {2\over 27}\right) M_N + {2\over 27}\left(28\,\sigma_{sN} + \sigma_{\pi N}\right)\right]\rho~~.
\label{eq:deltarho2}
\end{align}
Note again that the partial cancellation in the factor multiplying the nucleon mass $M_N$ on the right-hand side of this equation enhances the sensitivity of the (observable) quantity $\Delta(\rho=0) - \Delta(\rho)$ with respect to $\sigma_{sN}$.

Consider now first the limiting case with no change of the first spectral moment ($\Delta(\rho) = \Delta(0)$) when going from vacuum to nuclear matter (i.e. in simple terms: no in-medium mass shift of the $\phi$ meson). In this case with $\delta c_4 = 0$, we have the constraint
\begin{align}
\sigma_{sN} = {M_0\over 28} - {27\over 56} A_2^s\,M_N~~,
\label{eq:sigmasN}
\end{align}
with $M_0 = M_N - \sigma_{\pi N}$, the nucleon mass in the chiral limit. Choosing $M_0\simeq 890$ MeV and $A_2^s \simeq 0.04\pm 0.01$ from Table\,\ref{tab:parameters.matter}, one finds $\sigma_{sN} \simeq 12$ MeV with an uncertainty of about $25 \%$. The corresponding scalar strangeness content of the nucleon,
\begin{align}
y = {\langle N | \bar{s} s | N\rangle \over \langle N | \bar{q} q | N\rangle} = {\bar{m}_q\over m_s}\,{\sigma_{sN}\over \sigma_{\pi N}} ~~,
\label{eq:scontent}
\end{align}
using a quark mass ratio $\bar{m}_q / m_s = (m_u + m_d)/2 m_s \simeq 0.036$ \cite{Olive} and $\sigma_{sN} / \sigma_{\pi N} \simeq 1/4$, gives $y \sim 0.01$ in this limit. This is the baseline for orientation when we now proceed to a more systematic evaluation.

\begin{figure*}
\begin{center}
\includegraphics[width=9cm]{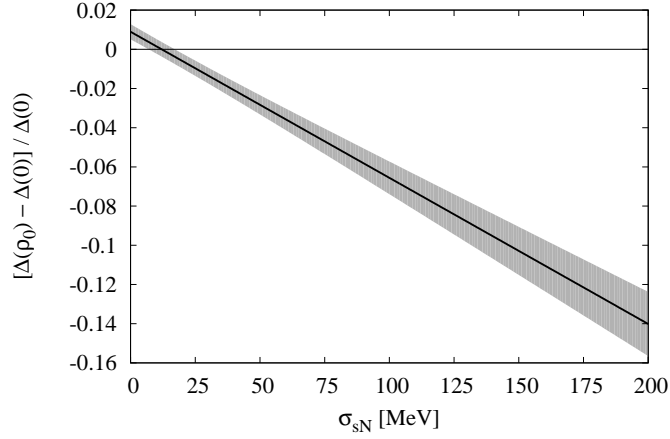}
\end{center}
\vspace{-0.6cm}
\caption{Relative change at normal nuclear matter density $\rho_0$ of the first moment of the $\phi$ meson spectral function as given by Eqs.\,(\ref{eq:deltarho2},\ref{eq:ratio}), as a function of the strangeness sigma term $\sigma_{sN}$ of the nucleon. The grey band reflects uncertainties of input parameters in $c_4$ and $\delta c_4$.}
\label{fig:mass.sigma}
\end{figure*}

A useful quantity to work with is the ratio
\begin{align}
{\Delta(\rho) - \Delta(0)\over \Delta(0)} = {\delta c_4(\rho)\over c_4}~~,
\label{eq:ratio}
\end{align}
which features an approximate measure of the in-medium mass shift $\delta m_{\phi}(\rho)$ of the $\phi$  meson: 
\begin{align}
{\Delta(\rho)\over \Delta(0)} -1 \sim {m^2_\phi(\rho)\over m^2_\phi} - 1\simeq {2\delta m_\phi(\rho)\over m_\phi}~. \nonumber
\end{align}
Computing $\Delta(0)$ and $\Delta(\rho)$ with our previously determined vacuum and in-medium spectral functions, respectively, we can plot the relative change of the first spectral moment, 
Eq.\,(\ref{eq:ratio}),
as a function of the strangeness sigma term $\sigma_{sN}$. At normal nuclear matter density $\rho = \rho_0 = 0.17$ fm$^{-3}$, such a plot is shown in Fig.\,\ref{fig:mass.sigma}. As as rough estimate, if the change of $\Delta(\rho)$ with respect to $\Delta(0)$ results from the in-medium downward shift of a narrow $\phi$ resonance, a negative mass shift of about 15 MeV at $\rho = \rho_0$ would 
constrain the strangeness sigma term at around $\sigma_{sN} \simeq 50$ MeV. 
A simple parametrization of the result in Fig.\,\ref{fig:mass.sigma} is given by 
\begin{align}
\frac{\Delta(\rho)}{\Delta(0)} \simeq 1.01 - C \sigma_{sN} \frac{\rho}{\rho_0}
\end{align}
with $C = 0.8 \pm 0.1\,\mathrm{GeV}^{-1}$. 

The $\sigma_{sN}$ dependence shown in Fig.\,\ref{fig:mass.sigma} displays the same qualitative trend as that found in \cite{Gubler}, in which the mass shift 
of the $\phi$ meson peak was obtained from the Borel sum rules of Eq.(\ref{eq:sumrule}) via a Bayesian extraction of the spectral function but without consideration of the 
in-medium $\phi$ width (for details of this 
approach, see also \cite{GublerOka}). For the same values of $\sigma_{sN}$, the masses reported in \cite{Gubler} turn out to be about $10$ MeV 
larger than what is found here. Part of this difference can be explained by our disregard of the twist-2 gluon condensate in $\delta c_4$. As can be 
understood from Eq.\,(11) in \cite{Gubler}, this term gives a contribution of $-\frac{7}{12} \frac{\alpha_s}{\pi} A^g_2 M_N \rho$ to $\delta c_4$\footnote{Note however that 
a term proportional to $\log(Q^2/\mu^2)$ in the Wilson coefficient of this operator has been neglected in \cite{Gubler} and the OPE of this term therefore still needs to be completed, 
for instance by making use of the method outlined in \cite{Zschocke2}. 
Our discussion about its effects can hence only be of qualitative nature.}, where $A^g_2$ is 
the first moment of the gluonic parton distribution and has been estimated as $A^g_2 = 0.36 \pm 0.15$ \cite{Gubler,Martin}, which leads to an increase of the mass by about 5 MeV. 
The remainder of the discrepancy is presumably due to the use of different sum rules (Borel sum rules vs. finite energy sum rules) and/or the different analysis methods. 

In the context of Fig.\,\ref{fig:mass.sigma}, let us furthermore briefly discuss most recent lattice results on $\sigma_{sN}$ of three different collaborations, 
which have appeared during the last few months \cite{DurrBMW,YangchiQCD,AbdelRehimETM} and which were all carried out at (or around) physical quark masses, 
therefore eliminating systematic uncertainties due to chiral extrapolations. 
Nevertheless, the results reported by the three collaborations still have a surprisingly wide spread: 
\begin{equation}
\begin{split}
\sigma_{sN} &= 105(41)(37)\, \mathrm{MeV}~~[52], \\
\sigma_{sN} &= 32.3(4.7)(4.9)\, \mathrm{MeV}~~[53], \\
\sigma_{sN} &=41.05(8.2)(^{+1.09}_{-0.69})\, \mathrm{MeV}~~[54].
\end{split}
\label{eq:sigmasN.lattice}
\end{equation} 
These uncertainties still prevent us so far from drawing a firm conclusion about the actual value of $\sigma_{sN}$. In combination with the sum rule results of this work, it is however possible to 
discuss the potential consequences of the findings of Eq.\,(\ref{eq:sigmasN.lattice}). For the largest value of \cite{DurrBMW}, the mass shift $\delta m_\phi(\rho_0)/m_\phi$ turns 
out to be about -3.5\,\%, which happens to coincide with the experimental result of the E325 experiment at KEK \cite{Muto}, where a negative mass shift of $35\pm7\,\mathrm{MeV}$ was 
extracted at $\rho = \rho_0$, although with a simple Breit-Wigner parametrization that is not expected to be reliable.  
On the other hand, the smaller values of the other two lattice QCD computations \cite{YangchiQCD,AbdelRehimETM} would rather point to much smaller mass shifts of $-1.0\,\%$ or less. 
It will be interesting to see whether such findings can be confirmed by the E16 experiment at J-PARC, which plans to measure the $\phi$ meson spectrum at nuclear matter density 
with significantly increased statistics compared to E325. 

\section{\label{Fourquark} Strange four-quark condensates in vacuum and nuclear matter}
Up to this point 
our FESR analysis has been restricted to the lowest two moments of $R_\phi(s)$. Using Eqs. (\ref{eq:sm1},\ref{eq:sm2}). 
the scale $s_0$ of Eq.\,(\ref{eq:ansatz}) has been determined consistently. 
The spectral functions $R(s)$ in vacuum and at finite density are fixed in this way. 
With this input we can now proceed one step further, 
turn to the next-higher moment and the corresponding sum rule of Eq.(\ref{eq:sm3}), and study the vacuum and in-medium four-quark condensates. Their commonly used parametrization is based on the factorization hypothesis: 
\begin{align}
\langle (\overline{s}\,\gamma_{\mu} \gamma_5,\lambda^a\,s)^2 \rangle\,  +\, \frac{2}{9} 
\langle (\overline{s}\,\gamma_{\mu} \,\lambda^a\, s) \sum_{q=u,d,s} (\overline{q}\,\gamma_{\mu} \,\lambda^a\, q)  \rangle 
= \, 
\frac{112}{81}\,\kappa_0 \,\langle \overline{s} s \rangle^2. 
\label{eq:fourquark1}
\end{align}
Here $\kappa_0 = 1$ stands for exact vacuum saturation. Any deviation from this value 
signals the degree of violation of the extreme factorization assumption. 
Such violations are due to intermediate states other than the vacuum that appear as part of a complete set of 
states inserted between the $\overline{s} s \,\overline{s} s$ or $\overline{s} s\, \overline{q} q$ operators. Among these, 
kaonic states are lowest in energy and are therefore expected to play an important role. They appear as intermediate states 
between Fierz transformed $\overline{s} s \,\overline{u} u$ or $\overline{s} s \,\overline{d} d$ operators.  
The factorization formula of Eq.\,(\ref{eq:fourquark1}) ignores such intermediate states and relegates their effects to an 
``effective" parameter $\kappa_0 > 1$. 
From these arguments one can already conjecture that $\kappa_0 = 1$ is not likely to be a realistic estimate. 

Generalizing the factorization hypothesis to 
finite density, the right-hand side of Eq.(\ref{eq:fourquark1}) changes to 
\begin{equation}
\frac{112}{81} \kappa_N(\rho)  \Big( \langle \overline{s} s \rangle^2 + 2 \frac{\sigma_{sN}}{m_s} \, \langle \overline{s} s \rangle \, \rho \Big), 
\label{eq:fourquark2}
\end{equation}
where $\kappa_N(\rho)$ generally depends on the density $\rho$ and the in-medium strange quark condensate has been expanded to leading order in $\rho$. 
Note that in the low-density limit, $\kappa_N$ and $\kappa_0$ are related as $\kappa_N(\rho \to 0) = \kappa_0$. 

Before proceeding with the discussion of the four-quark condensate terms and their role in the second moment sum rule, let us 
comment on other terms of dimension 6 that can appear in the OPE of the $\phi$ meson channel. 
First we point out that as long as $\kappa_0$ or $\kappa_N$ take values of order 1 and larger, the four-quark condensate term 
dominates over the other two terms of higher order in $m_s$ given in Eq.\,(\ref{eq:operesult3}). These terms can thus be safely neglected for the purposes of this discussion. 
Further terms that appear at dimension 6 in vacuum include the mixed condensate, $m_s \langle \overline{s} g \sigma_{\mu \nu} G^{\mu \nu} s \rangle$, 
and gluonic condensate involving three gluon fields, $\langle g^3 G^3 \rangle $. For the vector correlator, the Wilson coefficients of 
both these condensates are known to vanish at leading order in $\alpha_s$ \cite{Generalis} and are therefore suppressed. 
At finite density more condensates can emerge because of broken Lorentz symmetry. These involve expectation values of twist-2 or twist-4 operators which 
can be constructed from both quark and gluon fields. The OPE for the quark-operators was recently worked out in \cite{Gubler2}, the 
full result for the gluonic sector is however still missing (see \cite{Kim} for a discussion of such operators in the context of 
quarkonium sum rules). In this work, we will ignore all those non-scalar operators of dimension 6 in order to get a rough idea 
on the value of $\kappa_N$ at normal nuclear matter density. 
To reach a more quantitative conclusion, a full analysis taking into account all potentially large operators will however be required. 
We leave this issue to be studied in future work. 

Returning to Eqs.\,(\ref{eq:fourquark1}) and (\ref{eq:fourquark2}), we now 
combine our spectral function and the threshold parameters obtained from the zeroth and first moments with the FESR of Eq.\,(\ref{eq:sm3}) 
to extract possible constraints for $\kappa_0$ and $\kappa_N$. 
For simplicity we use here only the sum rules derived from the step-function ansatz of Eq.\,(\ref{eq:ansatz}). 
Specifically, we rewrite Eq.\,(\ref{eq:sm3}) to obtain an explicit expression for $\kappa_0$,  
\begin{equation}
\kappa_0 = -\frac{81}{224 \pi \,\alpha_s \langle \overline{s} s \rangle^2} \Bigg[\int_0^{s_0} ds \, s^2 R_{\phi}(s) - \frac{c_0}{3} s_0^3 \Bigg], 
\label{eq:kappa0}
\end{equation}
and for $\kappa_N$,   
\begin{equation}
\kappa_N = -\frac{81}{224 \pi \,\alpha_s \big( \langle \overline{s} s \rangle^2 + 2 \frac{\sigma_{sN}}{m_s} \, \langle \overline{s} s \rangle \, \rho \big)} 
\Bigg[\int_0^{s_0^{\ast}} ds \, s^2 R_{\phi}(s; \rho) - \frac{c_0}{3} s_0^{\ast\,3} \Bigg].  
\label{eq:kappaN}
\end{equation}

\begin{figure}
\begin{center}
\includegraphics[width=8cm]{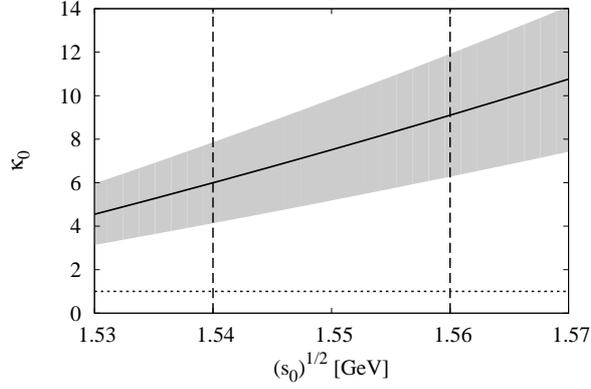}
\end{center}
\vspace{-0.6cm}
\caption{The factorization parameter $\kappa_0$ of the four-quark condensate defined in Eq.\,(\ref{eq:fourquark1}), 
extracted from the second moment sum rule (\ref{eq:sm3}), as a function of the scale $\sqrt{s_0}$ delineating low-energy and PQCD regions. 
The vertical dashed lines show the bounds of $\sqrt{s_0} = 1.55 \pm 0.01$, as obtained from the zeroth and first moment sum rules. 
The horizontal short-dashed line indicates $\kappa_0=1$, corresponding to the exact factorization hypothesis. 
}
\label{fig:four.quark}
\end{figure}

The resulting values of $\kappa_0$ are shown in Fig.\,\ref{fig:four.quark}.    
The uncertainties of $\kappa_0$ seen in this figure 
are related to cancellations that occur between the spectral function integral and the  
perturbative QCD term proportional to $s_0^3$. Combined with the errors attached to the strange quark condensate 
and to $\alpha_s$, the $\kappa_0$ values are constrained within a relatively broad window as indicated in Fig.\,\ref{fig:four.quark}. One observes a strong dependence on the 
scale parameter $\sqrt{s_0}$. A typical value,  $\sqrt{s_0} = 1.55$ GeV, gives  
\begin{equation}
\kappa_0 \sim 7 \pm 2. 
\end{equation}
Despite considerable uncertainties, one concludes at least qualitatively from this result that factorization based on vacuum dominance $(\kappa_0 = 1)$
is not expected to be a good approximation. 

\begin{figure}
\begin{center}
\includegraphics[width=7cm]{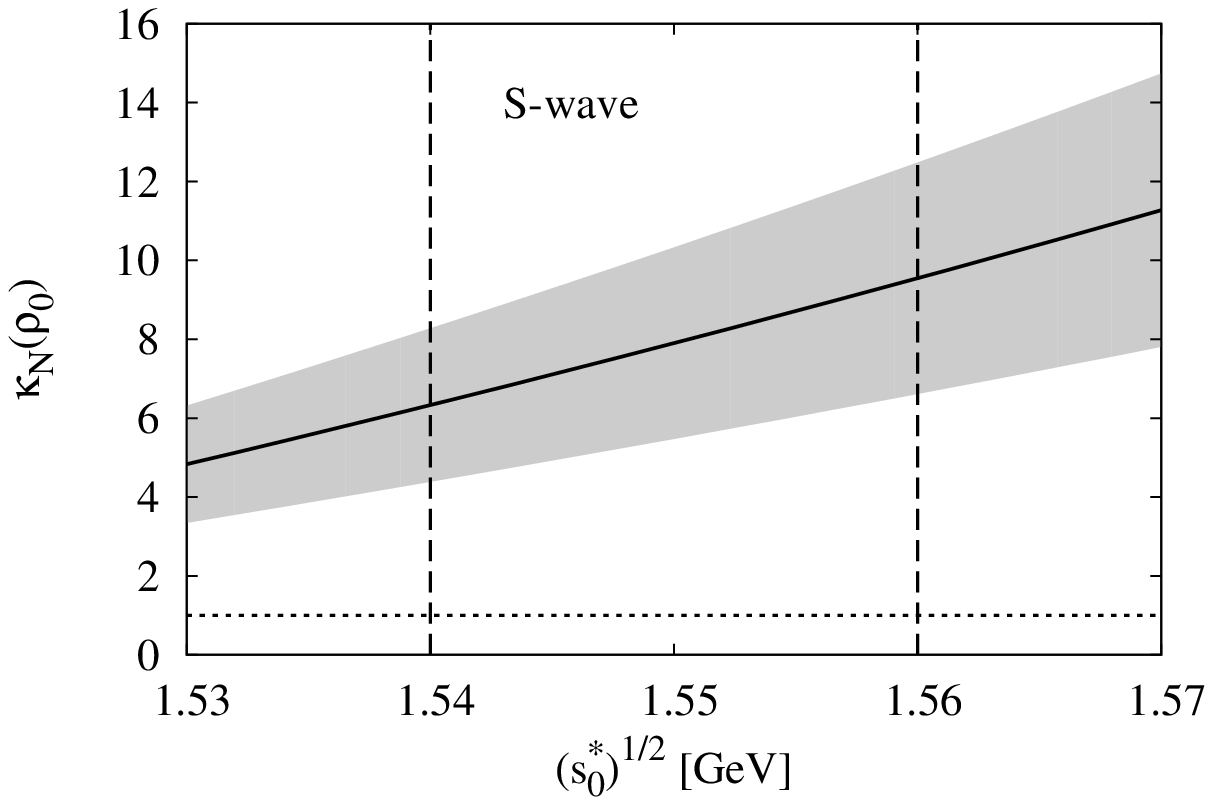}
\includegraphics[width=7cm]{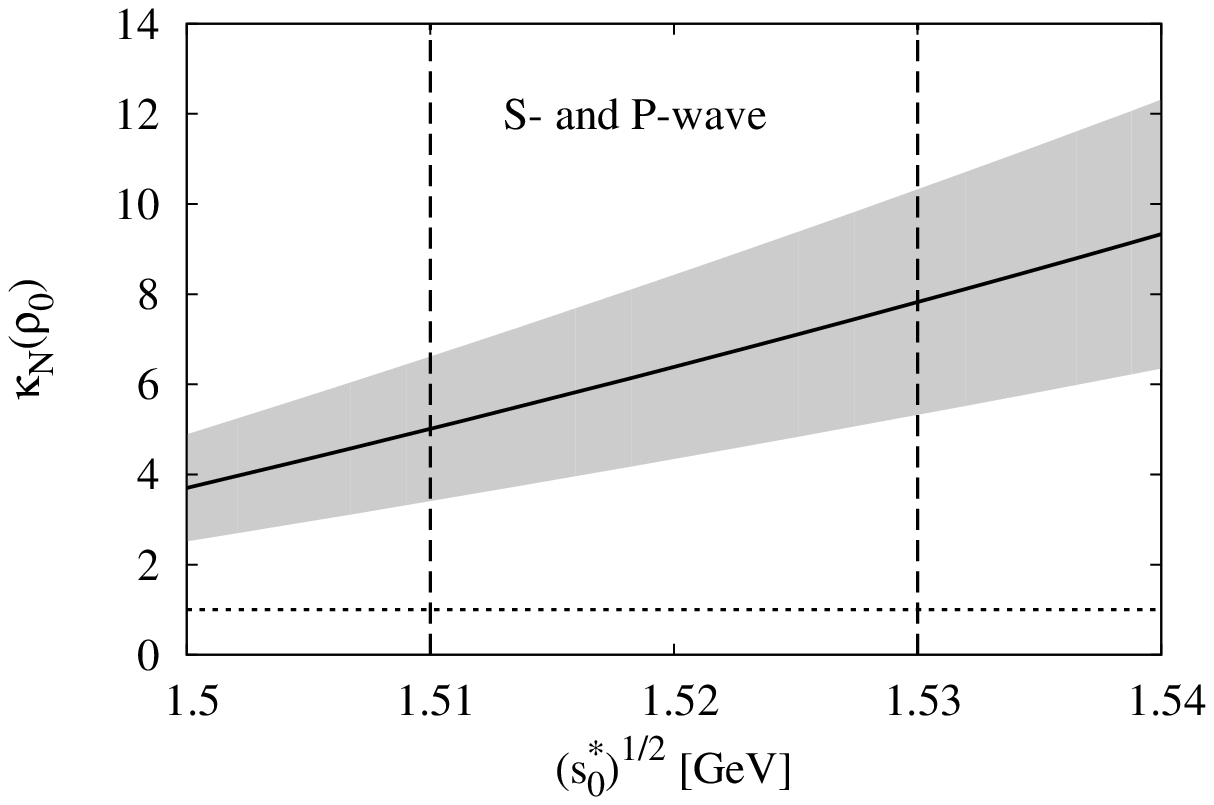}
\end{center}
\vspace{-0.6cm}
\caption{Same as Fig.\,\ref{fig:four.quark}, but for $\kappa_N$ at normal nuclear matter density $\rho_0$. In the 
left (right) plot, S-wave (S- and P-wave) $\overline{K}N$ interactions are included.}
\label{fig:kappa.matter}
\end{figure}

%
To study the behavior of the four-quark condensates at finite density the analysis is repeated using Eqs.\,(\ref{eq:fourquark2},\ref{eq:kappaN}) 
and the modified spectral functions of Fig.\,\ref{fig:vac}. 
The respective results are given in Fig.\,\ref{fig:kappa.matter}. 
For the S-wave case, the situation does not deviate much from the vacuum, 
the range for $\kappa_N$ at normal nuclear matter density being again 
\begin{equation}
\kappa_N \sim 7 \pm 2 \hspace{0.3cm} (\text{S-wave})  
\end{equation}
for $\sqrt{s_0} = 1.55$ GeV. 
As expected from the findings of the previous section, P-wave interactions have a somewhat stronger effect, 
leading to an overall moderate decrease of the allowed values of $\kappa_N(\rho_0)$ at $\sqrt{s_0} = 1.52$ GeV: 
\begin{equation}
\kappa_N \sim 6 \pm 2 \hspace{0.3cm} (\text{S- and P-wave}).  
\end{equation}
Even though these constraints are not very restrictive, they nevertheless indicate clearly
that the factorization hypothesis is expected to be strongly violated in nuclear matter, similar to what has been seen in the vacuum case. 
Furthermore, it is interesting to observe that the density dependence of $\kappa_N(\rho)$ appears to be 
rather weak. 

This is an issue that deserves a more detailed investigation, for instance by explicitly taking into account 
the intermediate single and multiple kaon states between the two-quark operators.  
We plan to tackle this problem in future work. 

\section{\label{Conclusion} Summary and Conclusions}
In this work we have studied the modifications of the $\phi$ meson spectral function in 
nuclear matter based on an effective field theory approach employing up-to-date $KN$ and $\overline{K}N$ interactions. 
The resulting spectra were then used to compute their zeroth and first moments, that are related to QCD parameters and 
condensates of mass dimension 2 and 4 via the finite energy sum rules (FESR). 

As a basis for our analysis at finite density, we have first constructed a spectral function for the $\phi$ meson in vacuum, based on 
a vector dominance model improved through the inclusion of kaon loops that couple to the bare $\phi$ meson. As seen in Fig.\,\ref{fig:vac.fit}, 
this allows for an excellent description of recent experimental $e^+ e^- \to K^+ K^-$ cross section data \cite{Lees} in the 
region of the $\phi$ resonance. To apply this spectrum to the analysis of moments, one however needs to model the onset of the 
continuum, which should approach the perturbative QCD limit at high energy. For this purpose, we have tested two different 
ans$\mathrm{\ddot{a}}$tze: the step function of Eq.\,(\ref{eq:ansatz}) and the schematic ramp function of 
Eqs.\,(\ref{eq:ansatz.ramp},\ref{eq:ansatz.ramp2}), motivated by the behavior of the channels involving one and two pions 
in addition to $\overline{K}K$ and shown 
in Fig.\,\ref{fig:vac.fit.2}. Combining these functional forms with the $\phi$ meson spectrum at low 
energy, our FESR analysis shows almost perfect consistency within errors for the 
characteristic scales that delineate the low and high energy ranges of the zeroth and first moment 
sum rules. 

Using this well constrained vacuum spectral function as a starting point, we next examined the effects of finite density.  
Assuming that the nucleons of the surrounding medium affect the $\phi$ meson primarily through interactions with kaons, 
density effects were taken into account by modifying the kaon propagators that enter into the kaon loops of the 
$\phi$ meson self-energy. To leading order in density, the modifications are determined by the free $KN$ and 
$\overline{K}N$ scattering amplitudes. 
To this end, we first employed S-wave $KN$ and $\overline{K}N$ amplitudes derived from 
chiral $SU(3)$ effective field theory, which are repulsive for the $KN$ and attractive and strongly 
energy dependent for the $\overline{K}N$ case. The latter property is related to the formation of the $\Lambda(1405)$ resonance 
in that channel. While for $KN$ it suffices to approximate the amplitude by a purely real constant, a more elaborate 
treatment is needed for $\overline{K}N$, for which we made use of a realistic coupled-channel approach based on a chiral $SU(3)$ 
effective Lagrangian. 
At normal nuclear matter density $\rho_0$, 
all this leads to a rather strong broadening of the $\phi$ meson peak ($\Gamma_{\phi} \simeq 24\,\mathrm{MeV}$) and only a small 
overall shift of its peak position. 
Including furthermore P-wave $KN$ and $\overline{K}N$ interactions involving $K$-hyperon intermediate 
states into the analysis, causes the broadening to increase further ($\Gamma_{\phi} \simeq 45\,\mathrm{MeV}$) and induces to 
a downward shift of the resonance maximum. 

Computing the first and zeroth moments of the finite density spectrum, the level of agreement among the FESR's can again be tested. 
While the sum rule for the zeroth moment [Eqs.(\ref{eq:sm1}) and (\ref{eq:sm1.ramp})] remains unaffected by finite density 
effects, the one for the first moment [Eqs.(\ref{eq:sm2}) and (\ref{eq:sm2.ramp})] receives corrections due to the modifications 
of scalar condensates and the emergence of new non-scalar condensates. The consistency of the two sum rules however 
turns out to be just as good as for the vacuum, unless the slope parameter of the ``ramp" function interpolating between low- and 
high-energy regimes becomes unphysically small. 

A special feature of the in-medium $\phi$ meson spectral function is the strong sensitivity of its first moment with respect to $\sigma_{sN}$, 
the strangeness sigma term of the nucleon, i.e. the degree to which the scalar density of strange quarks contributes to the nucleon ground 
state. At leading order in density, a linear relationship emerges between this sigma term and the in-medium shift of the first spectral 
moment (with the perturbative QCD limit subtracted). A measurement of this in-medium spectral moment of the $\phi$ would 
therefore add an empirical constraint to the ongoing discussion about $\sigma_{sN}$, recently promoted by several lattice QCD 
computations.  

Given the spectral function for both the vacuum and nuclear matter, it is possible to compute still higher moments and 
to study the respective sum rules. Here we have focused on the second moment. Through Eqs.(\ref{eq:sm3}) and 
(\ref{eq:sm3.ramp}) it is related to condensates of mass dimension 6, with the dominant contribution coming from four-quark 
condensates involving one or two strange quark-antiquark pairs. This sum rule can therefore be used to derive constraints on the 
values of these four-quark condensates. 
It is found that the frequently employed factorization hypothesis, assuming ground state dominance, is likely to be strongly violated 
both in vacuum and nuclear matter. 

Comparing our findings with those of earlier works, it is seen that our spectral functions have qualitatively similar features 
as those of previous studies \cite{Klingl,Klingl3,Cabrera}, namely relatively strong broadening and only a small negative mass shift of the $\phi$ meson peak. On the 
other hand, the sum rule input has changed significantly compared to \cite{Hatsuda,Zschocke} because of the possibility of much smaller $\sigma_{sN}$ values, 
which lead to reduced mass shift predictions. 

As a last point, let us discuss implications of the present study for future measurements of the in-medium $\phi$ meson spectral function, in 
particular with reference to the E16 experiment planned at J-PARC \cite{Kawama,Aoki}.  
In the present work the $\phi$ meson was considered to be at rest with respect to the surrounding medium. A direct comparison with experimental 
spectra corresponding to a $\phi$ in motion should therefore be made with caution. Earlier studies have already examined this issue 
\cite{Lee} to some extent, but more work is needed to fully understand the finite momentum effects. 
There are nevertheless a number of general conclusions that can be drawn from the present work. 
The in-medium spectra shown in 
Fig.\,\ref{fig:vac} exhibit a characteristically non-symmetric behavior around the $\phi$ resonance, with an enhancement of 
strength in the low-energy sub-resonance region caused by strong broadening and the opening of new decay channels. The possibility of such 
behavior should be taken into consideration in future analyses of experimental data. Furthermore, we stress again the usefulness of the two 
lowest spectral moments as they provide direct relations to the most relevant low-dimensional QCD condensates and their changes 
at finite density. High precision measurements of these and higher moments in nuclei would provide strong constraints for the behavior of 
various combinations of gluon and quark condensates in nuclear matter.

\section*{Acknowledgments}
This work is partially supported by BMBF grant 05P12WOCTB and by DFG through CRC 110 ``Symmetries and Emergence of Structure in QCD". 
The authors would like to thank Yoichi Ikeda for providing them with his code for computing the $\overline{K}N$ scattering amplitudes. 





\end{document}